\documentclass[a4paper,11pt]{article}

\usepackage{heppub}
\usepackage{hyperref}
\usepackage{graphicx}
\usepackage{amsmath}
\usepackage{amssymb}
\usepackage{mathtools}

\usepackage{xspace}
\usepackage[small]{subfigure}
\usepackage{slashed}
\usepackage{dsfont}
\usepackage[normalem]{ulem}
\usepackage{xcolor}
\usepackage{verbatim}
\usepackage{stmaryrd}
\newcommand{\bn}{{\bar n}}

\newcommand{\cP}{{\cal P}}
\def\bnslash{\bar n\!\!\!\slash}
\def\nslash{n\!\!\!\slash}

\newcommand{\SCETb}{\mbox{${\rm SCET}_{\rm II}$}\xspace}

\newcommand{\mb}[1]{\boldsymbol{#1}}
\newcommand{\eps}{\epsilon}
\newcommand{\im}{\mathrm{i}}
\newcommand{\qpsq}{\vec q_\perp^{\,\prime 2}}
\newcommand{\qsq}{\vec q_\perp^{\, 2}}
 
\newcommand{\cO}{\mathcal{O}}

\newcommand{\pslash}{p\!\!\!\slash}

\newcommand{\qslash}{q\!\!\!\slash}

\newcommand{\ellslash}{\ell\!\!\!\slash}

\newcommand{\psl}{p\!\!\!\slash}
\newcommand{\ra}{\rightarrow}

\newcommand{\as}{\alpha_s}
\newcommand{\bas}{\bar \alpha_s}

\newcommand{\nn}{\nonumber}
\newcommand{\nlp}{\mathrm{HP}}
\newcommand{\ps}{\text{pure-singlet}}
\newcommand{\df}{\mathrm{d}}

\newcommand{\cF}{\mathcal{F}}

\newcommand{\Kout}{P_{e}^{\prime}\,\!}
\newcommand{\Kin}{P_{e}\,\!}
\newcommand{\x}{x_{b}}

\def\dbar{{\:\mathchar'26\mkern-12mu d}}

\usepackage{todonotes}
\allowdisplaybreaks[4]

\DeclareRobustCommand{\Ref}[1]{Ref.~\cite{#1}}

\preprint{
\begin{flushright}
MIT-CTP/5529\\
DESY-23-024\\
UWThPh 2023-3
\end{flushright}
}

\title{Small-$x$ Factorization from Effective Field Theory}

\author[a,b]{Duff Neill,}
\affiliation[a]{Theoretical Division, MS B283, Los Alamos National Laboratory, Los Alamos, NM 87545, USA}
\affiliation[b]{Center for Theoretical Physics, Massachusetts Institute of Technology, Cambridge, MA~02139, U.S.A.}

\author[b,c,d,e]{Aditya Pathak,}
\affiliation[c]{Deutsches Elektronen-Synchrotron DESY, Notkestr. 85, 22607 Hamburg, Germany}
\affiliation[d]{University of Manchester, School of Physics and Astronomy, Manchester, M13 9PL, United Kingdom}
\affiliation[e]{University of Vienna, Faculty of Physics, Boltzmanngasse 5, A-1090 Vienna, Austria}

\author[b]{Iain W.~Stewart}

\emailAdd{duff.neill@gmail.com}
\emailAdd{aditya.pathak@desy.de}
\emailAdd{iains@mit.edu}

\abstract{
We derive a factorization theorem that allows for resummation of small-$x$ logarithms by exploiting Glauber operators in the soft collinear effective field theory. Our analysis is carried out for the hadronic tensor $W^{\mu\nu}$ in deep inelastic scattering, and leads to the definition of a new gauge invariant soft function $S^{\mu\nu}$ that describes quark and gluon emission in the central region.  This soft function provides a new framework for extending resummed calculations for coefficient functions to higher logarithmic orders. Our factorization also defines impact factors by universal collinear functions that are process independent, for instance being identical in small-$x$ DIS and Drell-Yan.
}
\keywords{QCD, BFKL, Colliders, Renormalization Group}

\begin{document}
	
\maketitle

\pagebreak

%%%%%%%%%%%%%%%%%%%%%%%%%%%%%%%%%%%%%%%%%%%%%%%%%%%%%%%%%%%%%%%%%%%%%%%%%%%%%%%%
\section{Introduction}
\label{sec:Intro}
%%%%%%%%%%%%%%%%%%%%%%%%%%%%%%%%%%%%%%%%%%%%%%%%%%%%%%%%%%%%%%%%%%%%%%%%%%%%%%%%

The forward scattering regime of a field theory has long been an object of interest. Of chief concern is the question of what form should a strongly-interacting bound state take, such that when scattering at high-energies, the interactions of its constituents with the other projectile correctly unitarizes the cross-section \cite{Gribov:1983ivg}. Typically, one tries to reduce questions about the structure of the bound state in high energy scattering by relating it to universal correlation functions, like the parton distribution function, through the process of collinear factorization (or equivalently, the twist expansion). This requires a hard momentum transfer to occur in the scattering process at a scale $Q^2\gg \Lambda_{\rm QCD}^2$. Then through asymptotic freedom, one can access the partonic constituents of the bound state. The high-energy scattering regime is accessed when we keep the momentum transfer $Q^2$ between the scattering states fixed, and instead increase the center-of-mass energy $s$ to asymptotically large scales. Historically in quantum chromodynamics, two general approaches to the high-energy scattering regime have evolved, both united by the critical role played within each by the so-called Balitsky-Fadin-Kuraev-Lipatov (BFKL) equation~\cite{Lipatov:1976zz,Kuraev:1976ge,Kuraev:1977fs, Balitsky:1978ic}. One is dominated by concerns of how the bound state manages to unitarize the scattering cross-section through the formation of a saturated state\footnote{A leading contender for this saturated state in nuclei is given by the color-glass-condensate~\cite{McLerran:1993ka,McLerran:1993ni,McLerran:1994vd}. } of partonic constituents, created in a dynamical process described through a hierarchy of nonlinear functional equations known as the ``B-JIMWLK'' equations, or its functional closure through the large $N_c$ limit into the BK equation \cite{Balitsky:1995ub,Kovchegov:1999yj,JalilianMarian:1996xn,JalilianMarian:1997gr,Iancu:2001ad}. 
The second approach is concerned with how to connect the resummation carried out by the BFKL equation to the DGLAP resummation appearing within collinear factorization with parton distribution functions, since within collinear factorization we have the most reliable perturbative control. This approach was initiated by Ref.~\cite{Catani:1994sq}, that first resummed the leading small-$\x$ logarithms in the coefficient functions and anomalous dimensions defined through the standard dimensional regularization approach to collinear factorization, and sparked several groups to address concerns about the stability of the perturbation series (Refs.~\cite{Ciafaloni:1998iv,Salam:1998tj,Ciafaloni:1999yw,Ciafaloni:2003rd,Ciafaloni:2003kd,Ciafaloni:2003ek} and  \cite{Altarelli:1999vw,Altarelli:2001ji,Altarelli:2003hk,Altarelli:2005ni,Marzani:2007gk} and \cite{Thorne:2001nr}) when extending the BFKL equation to next-to-leading order, Ref. \cite{Fadin:1998py}. In this paper we bring a new set of tools to these problems using soft-collinear effective field theory (SCET)~\cite{Bauer:2000ew,Bauer:2000yr,Bauer:2001ct,Bauer:2001yt,Bauer:2002nz}.
We will make use of the SCET based description of the forward scattering limit with Glauber operators from~\Ref{Rothstein:2016bsq}.

An important classic observable for studying the forward scattering regime in QCD is Deep Inelastic Scattering (DIS), $e^-p\to e^- X$. The scattering cross-section directly probes the structure of the proton, which can be summarized by the so-called structure functions, $F_p(\x,Q^2)$, where $Q^2$ is the momentum transfered into the strongly-interacting or hadronic sector, and $\x$ in the collinear factorization regime has the interpretation of the momentum fraction of a parton within the proton, projected along the direction of the proton. As $\x\rightarrow 0$, we naturally probe the forward scattering region, as we describe in detail later on.

A goal of this work is to setup a formalism involving factorization of small-$\x$ momentum regions with gauge-invariant objects with manifest power-counting, to lay ground work for resummation at higher orders and for other processes than inclusive DIS. This can be naturally achieved in the effective theory framework, which we demonstrate by
deriving a small-$x$ factorization theorem for the $W^{\mu\nu}$ hadronic tensor in DIS using Glauber SCET, 
\begin{align}\label{eq:fact}
   W^{\mu\nu}(q,P) = \int d^{d-2}k_\perp 
   S^{\mu\nu}\Big(q,k_\perp,\frac{\nu}{xP^-},\epsilon\Big) 
   \, C\Big(k_\perp,P,\frac{\nu}{P^-},\epsilon\Big) \,.
\end{align}
Here $S^{\mu\nu}$ is a soft function and $C$ a collinear function and renormalization group evolution in $\nu$ sums small-$x$ logarithms.
The soft function $S^{\mu\nu}$ is analogous to the dipole function defined in the B-JIMWLK framework. Likewise, the collinear function plays a role similar to impact factors in the small-$x$ literature   and incorporates higher order corrections related to the initial hadronic state.
This result is valid at next-to-leading-logarithmic (NLL) order, and enables the computation of DIS coefficient functions at this order.
This also enables the universality of functions to be made manifest with matrix elements of definite operators that appear in multiple processes. 
While our formulation shares the goals of earlier small-$\x$ resummation approaches, it differs in some key ways.
In the approach of Refs.~\cite{Catani:1994sq,Ciafaloni:1998iv,Salam:1998tj,Ciafaloni:1999yw,Ciafaloni:2003rd,Ciafaloni:2003kd,Ciafaloni:2003ek,Altarelli:1999vw,Altarelli:2001ji,Altarelli:2003hk,Altarelli:2005ni,Marzani:2007gk} 
different off-shell cross sections for each individual quark/gluon channel and structure functions were required. In contrast, in \eq{fact} a single gauge invariant soft function $S^{\mu\nu}$ replaces these offshell cross sections and captures the process dependence. 
In addition, our collinear function is defined by an operator matrix element with hadron states and is universal across processes from the start, depending only on the properties and momentum of the hadron being probed.  In particular, the factorization for Drell-Yan will involve the same collinear function with proton matrix elements as in DIS, and have a different process-dependent soft function.  The simplicity of this universality can be contrasted with other approaches, such as Ref.~\cite{Ciafaloni:1998hu}, 
where maintaining process independence while incorporating higher order corrections related to the initial state is more challenging.

In \secn{review} we review the small-$x$ DIS kinematics, structure function decomposition, and Glauber SCET Lagrangian, and discuss the structure of small-$x$ logarithms in DIS and previous methods used to resum them. 
In \secn{modes} we determine the soft and collinear modes, power counting, and operators needed for our EFT description of small-$x$ DIS.
In \secn{factorization} we derive the small-$x$ factorization theorem for $W^{\mu\nu}$ in \eq{fact}. 
In \secn{fixedorder} we carry out fixed order calculations of the soft and collinear functions. Our leading order calculations of $S^{\mu\nu}$ for the $F_2$ and $F_L$ structure functions, directly verifies that this function provides a proper replacement of the off-shell cross sections used in earlier approaches. Our next-to-leading order (NLO) calculation of $C$ enables us to directly verify that it satisfies the $d$-dimensional BFKL equation. 
In \secn{resummation} we use our setup to reproduce the known LL resummation of the hard coefficient functions in the twist expansion, which enables us to highlight differences in the intermediate steps.  
In \secn{conclusion} we conclude and 
briefly mention the additional ingredients needed to extend the small-$\x$ resummation of coefficient function and anomalous dimension to NLL with our framework.

%%%%%%%%%%%%%%%%%%%%%%%%%%%%%%%%%%%%%%%%%%%%%%%%%%%%%%%%%%%%%%%%%%%%%%%%%%%%%%%
\section{Small-$x$ Logarithms in DIS}
\label{sec:review}
%%%%%%%%%%%%%%%%%%%%%%%%%%%%%%%%%%%%%%%%%%%%%%%%%%%%%%%%%%%%%%%%%%%%%%%%%%%%%%%%

For simplicity, here we consider the case of DIS with unpolarized proton struck with a virtual photon. In \secn{kin} we review the DIS variables and kinematics, and in \secn{Fp} the Lorentz decomposition of the cross section in terms of structure functions.  In \secn{logx} we  review  the set of terms in the perturbative expansion of hard-collinear twist-2 coefficient functions and anomalous dimensions that are enhanced in the small-$\x$ region. Finally in \secn{CH} we briefly describe the small-$\x$ and DGLAP resummation 
achieved for both anomalous dimensions and coefficient functions at LL accuracy by Catani and Hautmann in \Ref{Catani:1994sq}.
	
%%%%%%%%%%%%%%%%%%%%%%%%%%%%%%%%%%%%%%%%%%%%%%%%%%%%%%%%%%%%%%%%%%%%%%%%%%%%%%%%
\subsection{Kinematics}

\label{sec:kin}
We define the Lorentz vector momentum variables: $\Kin$ and $\Kout$ as the initial and final electron momenta, $P$ as the initial proton momentum, $P_X$ as the total momentum of the hadronic final state and $q$ as the momentum of the virtual photon. Thus total momentum conservation is $\Kin+P=\Kout +P_X$, and the Mandelstam variables are $s=(\Kin+P)^2$ and $t=q^2 = (\Kin-\Kout)^2$. The standard DIS kinematic variables $Q^2$, $\x$, and $y$ are then:
\begin{align}\label{eq:DISKin}
Q^2 = -q^2 > 0 \,, \qquad 
\x = \frac{Q^2}{2 P\cdot q} \,, \qquad 
y = \frac{P\cdot q}{P\cdot \Kin} = \frac{Q^2}{\x s}  \,,
\end{align}
where $0\le \x\le 1$ and $0\le y\le 1$. Equivalently, we may trade $y$ for the center of mass energy $s$, if $Q^2$ and $\x$ are fixed. For simplicity we work with a massless electron, and drop the proton mass $m_P^2 \ll Q^2$, which yields
\begin{align}\label{eq:DISMassless}
	Q^2 = - 2 \Kin \cdot q \, .
\end{align}
For our power counting and factorization analysis it is convenient to work with the center of mass frame of the $e^-$-proton collision. The choice of this frame makes it easier to demonstrate how the small-$\x$ limit is related to a forward scattering process. Here we have
\begin{align}
\Kin^\mu = \Kin^+ \, \frac{\bn^{\mu}}{2}  
      =   \sqrt{s}\, \frac{\bn^{\mu}}{2}
  \,, \qquad 
P^\mu = P^-\, \frac{n^{\mu}}{2}\label{eq:proton_cms}
      =  \sqrt{s}\, \frac{n^{\mu}}{2} \, ,
\end{align}
where $n=(1,0,0,1)$ and $\bar n=(1,0,0,-1)$ are light-like vectors along the electron and proton beam axes, respectively, and we drop both the electron and proton masses. We define plus and minus light-cone components for any vector $p^\mu$ by 
\begin{align}\label{eq:LCDef}
	p^\mu = p^+ \frac{\bn^\mu}{2} + p^- \frac{n^\mu}{2} + p_\perp^\mu \, , 
\end{align}
with $p^+=n\cdot p$ and $p^-=\bn\cdot p$. Here $p_\perp^\mu$ refers to the two components of $p^\mu$ that are orthogonal to both $n^\mu$ and $\bn^\mu$. The momentum squared is given by
\begin{align}
	p^2 = p^+ p^- + p_\perp^2  = p^+ p^- - \vec p_{\perp}^{\,2} \, ,
\end{align}
where the minus sign results from the metric signature for the transverse coordinates. Using \eqs{DISKin}{DISMassless} and expressing the light-like vectors $n^\mu$ and $\bn^\mu$ in terms of the electron and proton momentum in \eq{proton_cms} we have
\begin{align}
	q^+  = \frac{2 P \cdot q }{\sqrt{s}} = \frac{Q^2}{x_b \sqrt{s}} \, ,\qquad 
	q^- = \frac{2 P_e \cdot q}{\sqrt{s}} = - \frac{Q^2}{\sqrt{s}} \, .
\end{align}
such that
\begin{align}\label{eq:qmu}
	q^\mu &= \frac{Q^2}{x_b \sqrt{s}} \frac{\bn^\mu}{2}  - \frac{Q^2}{\sqrt{s}} \frac{n^\mu}{2} + q_\perp^\mu
	= y \sqrt{s} \frac{\bn^\mu}{2} - x_by \sqrt{s} \frac{n^\mu}{2} + q_\perp^{\mu} 
  \,.
\end{align}
Furthermore, using  $Q^2 = - q^2$, we have
\begin{align}\label{eq:qperpSq}
	\vec q_{\perp}^{\,2} = Q^2 (1-y) \, .
\end{align}
The invariant mass of the hadronic final state is given by $P_X^2=(q+P)^2$, yielding
\begin{align}\label{eq:invariant_mass}
P_X^2 = Q^2 (1-\x)/\x. 
\end{align}

%%%%%%%%%%%%%%%%%%%%%%%%%%%%%%%%%%%%%%%%%%%%%%%%%%%%%%%%%%%%%%%%%%%%%%%%%%%%%%%%
\subsection{DIS Structure Functions}
\label{sec:Fp}

For QCD with $n_f$ flavors of massless quarks, the DIS cross section can be written as
\begin{align}
\frac{\df \sigma}{\df \x\,\df y}(e^-p\ra e^-X) &= \frac{2\pi y \alpha^2}{Q^4}%\sum_X (2\pi)^4 \delta^4(P+q-P_X ) |\langle {\cal{M}} \rangle |^2 \nn \\&=
L_{\mu \nu } (\Kin,q) W^{ \mu \nu}(P,q) \, .
\end{align}
The leptonic tensor, assuming only photon exchange for unpolarized scattering and a massless electron, is given by
\begin{align}
	L^{\mu\nu} = 2 \big(P_e^\mu P_{e}^{\prime \nu} + P_e^\nu P_{e}^{\prime \mu} - g^{\mu\nu} P_e \cdot P_e^\prime \big) \, ,
\end{align}
and the hadronic tensor has the definition
\begin{align}
\label{eq:WmunuDef}
W^{\mu \nu}  &= \frac{1}{4\pi}\sum_X \langle P | J^{\mu \dagger}(0) | X \rangle \langle X | J^{\nu}(0) | P \rangle (2\pi)^4 \delta^4(P + q - P_X) \, \nn \\
&=\frac{1}{4\pi}\sum_X \int \df^4z\langle P | J^{\mu \dagger}(0) e^{\im z\cdot (P + q - P_X) }| X \rangle \langle X | J^{\nu}(0) | P \rangle (2\pi)^4 
\nn\\
&=\frac{1}{4\pi}\sum_X \int \df^4ze^{\im z\cdot q  }\langle P | e^{\im z\cdot\mathbb{P}}J^{\mu \dagger}(0) e^{-\im z\cdot\mathbb{P}}| X \rangle \langle X | J^{\nu}(0) | P \rangle \nn\\
&=\frac{1}{4\pi} \int \df^4ze^{\im z\cdot q  }\langle P | J^{\mu \dagger}(z) J^{\nu}(0) | P \rangle \,. 
\end{align}
We have eliminated the sum over the final state, using the translational invariance of the theory to remove the momentum conserving delta function. The resulting sum over a complete set of states is now the identity. The position space operator $J^{\mu}$ refers to the strongly interacting contribution to the electromagnetic current with which we are probing the proton's structure. The variable $X$ refers to the hadronic final state, where we sum inclusively over all possible configurations. The leptonic tensor is easily computed at tree-level. For unpolarized $e^-p$ scattering, the hadronic tensor $W^{\mu\nu}$ can be decomposed in terms of structure functions $F_2(x,Q^2)$ and $F_L(x,Q^2)$ via~\cite{Blumlein:2012bf,Manohar:1992tz},
\begin{align}\label{eq:F_2_F_L}
  W^{\mu\nu} = \Big( - g^{\mu\nu} + \frac{q^\mu q^\nu}{q^2}\Big) F_1 - \Big(q^\mu +2\x P^\mu\Big)
  \Big(q^\nu +2\x P^\nu \Big)  \frac{F_2}{2\x q^2} \,,
\end{align} 
where
\begin{align}
    F_L (x_b,Q^2) = F_2 (x_b,Q^2) - 2x F_1 (x_b,Q^2) \, .
\end{align}
We can restrict to these two terms, since we are only considering vector probes of QCD, and not axial vector, and exploit the  $P$-$C$-$T$ symmetries of QCD.
 
The inclusive factorization theorem for the DIS structure functions expresses them as a convolution between partonic cross sections $H_{2}^{(\kappa)}$ and $H_{L}^{(\kappa)}$ (also referred to as inclusive coefficient functions), and parton distribution functions $f_{\kappa/p}$ for a parton $\kappa$ in the proton $p$,
\begin{align}\label{eq:structure_f_factorization}
  \frac{1}{\x}F_2(\x,Q^2) &= \sum_{\kappa} \int_{\x}^1\! \frac{\df \xi}{\xi} \: H_{2}^{(\kappa)}\Big(\frac{\x}{\xi},Q,\mu\Big) f_{\kappa/p}(\xi,\mu)
   +{\cal O}\Big(\Lambda_{\rm QCD}^2/Q^2 \Big) \,,\nn\\
  \frac{1}{\x}F_L(\x,Q^2) &= \sum_{\kappa} \int_{\x}^1\! \frac{\df \xi}{\xi} \: H_{L}^{(\kappa )}\Big(\frac{\x}{\xi},Q,\mu\Big) f_{\kappa/p}(\xi,\mu)
   +{\cal O}\Big(\Lambda_{\rm QCD}^2/Q^2 \Big) \,.
\end{align}
The DGLAP evolution in $\mu$ sums logarithms between the scale $Q$ in $H_{2,L}^{(\kappa)}$ and the scale $\Lambda_{\rm QCD}$ intrinsic to the parton distrbution functions $f_{\kappa/p}$. In writing Eq. \eqref{eq:structure_f_factorization}, we have adopted the standard conventions of Refs.~\cite{Moch:1999eb,Vermaseren:2005qc}. 
Note that writing the integral for the factorization theorem in terms of the ratios $x_b/\xi$ in $H_{2,L}^{(\kappa)}$ and $\df\xi/\xi$, yields the factors of $1/x_b$ on the left hand side.
Following standard terminology, we will often call the factorization with PDFs as the twist expansion~\cite{PhysRevLett.22.744,Gross:1971wn,Brandt:1970kg,Christ:1972ms,Georgi:1974wnj}, most often defined by the naive scaling dimension minus spin for operators in the OPE.
One can construct projectors from the external momenta of the problem to isolate the appropriate structure function one is interested in. These are given in  Refs.~\cite{Moch:1999eb,Vermaseren:2005qc}:
\begin{align} \label{eq:FL}  
    \frac{1}{\x}F_2 (\x,Q^2) &=-\frac{2}{2-2\epsilon}\Big((3-2\epsilon)\frac{q^2}{(q\cdot P)^2}P_{\mu}P_{\nu}+g_{\mu\nu}\Big)  W^{\mu\nu} 
\,, \nn\\
 \frac{1}{\x}F_L(\x,Q^2) &=-\frac{2q^2}{(P\cdot q)^2}P_{\mu}P_{\nu}  W^{\mu\nu} \, .
\end{align}
However, for calculations at leading power in the forward scattering limit, we will find that a projector based on the electron's initial momentum is more natural for $F_2$:
\begin{align}\label{eq:F2withElectron}
    \frac{1}{\x}F_2 (\x,Q^2) &=-\frac{2(P_{e}\cdot q)^2}{s\Big(P_{e}\cdot q\, P\cdot q - s q^2/4\Big)}\Big(P_{\mu}P_{\nu}-\frac{(P\cdot q)^2}{(P_{e}\cdot q)^2}P_{e\mu}P_{e\nu}\Big)  W^{\mu\nu} \, .
\end{align}
In terms of the light-cone coordinates given in \secn{kin}, in the $e^-$-proton CM frame the projectors are given by
\begin{align}\label{eq:F2FLExp}
	\frac{1}{\x}F_2(\x, Q^2) = \frac{(n\cdot q)^2}{2\qsq} 
	\big(\bn_\mu \bn_\nu - x^2 n^\mu n^\nu\big) W^{\mu\nu} 
	\,, \quad
	\frac{1}{\x}F_L(\x,Q^2) = \frac{2Q^2}{(n\cdot q)^2} n_\mu n_\nu W^{\mu\nu} \, .
\end{align}

\subsection{Small $\x$ Logarithms in DIS}
\label{sec:logx}

A primary goal of factorization in the forward scattering limit is to resum the perturbative expansion of the structure functions. Typically, the structure functions are calculated in perturbation theory using dimensional regularization, since the functions themselves are infra-red divergent. These infra-red divergences may be absorbed into the parton distribution functions using the factorization formula of \eq{structure_f_factorization}. The standard definition of the Mellin moment of the structure functions is given as:
\begin{align}\label{eq:Mellin}
	\bar F_p(N,Q^2)=\int_{0}^{1}\frac{dx}{x}x^{N}\Big(\frac{1}{x}F_p(x)\Big)\,.
\end{align}  
Then when calculating the structure function in perturbation theory using leading twist-contribution, the result takes the form in Mellin space:
\begin{align}\label{eq:pert_structure_f_factorization}
	\bar F_p^{(\kappa )}(N)&=	\sum_{\kappa'} \bar H_{p}^{(\kappa')}\Big(N,\frac{Q^2}{\mu^2},\alpha_s(\mu^2)\Big)  
	\bar \Gamma_{\kappa'\kappa}	\big(\alpha_s(\mu^2), N, \eps\big)
	\, ,
\end{align}
where
\begin{align}\label{eq:TransitionFunction}
	\bar \Gamma_{\kappa'\kappa}	\big(\alpha_s(\mu^2), N, \eps\big) \equiv {\rm P} \, \exp \Bigg(\int_{0}^{\alpha_s(\mu^2)}\frac{d\alpha}{\beta(\epsilon,\alpha)}\mb{\gamma^s}(\alpha,N)\Bigg)_{\kappa' \kappa } 
  \,.
\end{align}
The finite remainder $\bar H_{P}^{(\kappa)}$ is the renormalized coefficient function of \eq{structure_f_factorization} when using the $\overline{\rm MS}$ scheme for the PDFs. 
The subscript $p=2,L$ selects for the corresponding tensor structure for each structure function, and the indicies $\kappa, \kappa'$ denote the partonic states used for the perturbative calculation. In particular, $\kappa$ labels the partonic state that we use to evaluate the structure function of \eq{WmunuDef}. For physical applications one wants to calculate the structure function using nucleon states, but within perturbation theory one has only access to perturbative states of quarks and gluons. To extract the coefficient functions it suffices to use initial quark and gluon states, since these coefficients are state independent. When using these partonic states we label the perturbative calculation of each structure function with index $\kappa$ to indicate the initial partonic state.  From hereon we will only consider the singlet contribution of quark and anti-quark as the non-singlet channels are power suppressed for small-$x$.
Thus $\kappa,\kappa' = q,g$ for each parton flavor, where $q$ refers to the quark/anti-quark contribution to the singlet parton density. 
The beta-function in dimensional regularization is
\begin{align}
 \beta(\epsilon,\alpha_s)\equiv \mu^2 \frac{\df}{\df\mu^2}\, \alpha_s = -\epsilon\,\alpha_s +\beta(\alpha_s)
 \,,
\end{align}
and $\mb{\gamma^s}(\alpha,n)$ in \eq{pert_structure_f_factorization} corresponds to the space-like DGLAP anomalous dimension matrix, or the anomalous dimensions of twist-two local operators. 
These anomalous dimensions control the logarithmic structure with respect to $Q^2$, the invariant mass of the photon probe of the nucleon's state.  
In \eq{pert_structure_f_factorization} the exponentiated anomalous dimension, $\bar \Gamma_{\kappa'\kappa}$, (with path-ordering for the matrix products) factors out all IR divergences of the perturbative calculation, corresponding precisely with the partonic PDFs. 
This occurs because in the $\overline{\rm MS}$ scheme the bare partonic PDFs are scaleless, so after UV renormalization they are purely determined by a series of infrared $1/\epsilon$ poles, controlled by the UV anomalous dimension.
After these IR divergences are extracted from the perturbative calculation via \eqs{pert_structure_f_factorization}{TransitionFunction}, we can then identify the remaining piece in \eq{structure_f_factorization} with the IR finite coefficient functions $\bar H^{(\kappa)}_p$ in \eq{pert_structure_f_factorization}.
The coefficient function and the anomalous dimensions have the perturbative expansion,
 \begin{align}\label{eq:Hexpansion}
 	\bar H_p^{(\kappa)}\bigg(N, \frac{Q^2}{\mu^2},\as(\mu^2)\bigg)	&=\bigg(\frac{1}{n_f}\sum_{i = 1}^{n_f}e_{q_i}^2\bigg)	\Bigg[ 2n_f \delta_{p,2}\delta_{\kappa,q}	+ \sum_{\ell=1}^\infty \tilde h_{p,\ell}^{(\kappa)}	\Big(\frac{Q^2}{\mu^2},N\Big)\bigg(\frac{\as(\mu^2)}{\pi}\bigg)^{\ell}	\Bigg] \,	,		\\
 	\boldsymbol{\gamma^s}  \big(N,\as(\mu^2)\big)	&=	\sum_{\ell = 1}^\infty \mb{\gamma^{s(\ell)}}	(N)\Big(\frac{\as}{\pi}\Big)^\ell \,	.	\nn
 \end{align}
The first term in the expansion of $\bar H_p^{(\kappa)}$ indicates that for $p = 2$ and the quark channel, we have the leading contribution to the $\bar F_2^q$ structure function from direct coupling of photons to quarks. For simplicity, below we will drop the average electromagnetic charge prefactor appearing in \eq{Hexpansion}.

Now, under this Mellin convention, the logarithms of $\x$ appear as poles located at $N=1$:
\begin{align}\label{eq:MellinPole}
	\int_{0}^{1}\frac{dx}{x} x^{N}\Big(\frac{1}{x}\text{ln}^{\ell-1}\big(x\big)\Big)&=(-1)^{\ell}\frac{\Gamma(\ell)}{(N-1)^{\ell}}\,.
\end{align}
In what follows, we will find it convenient to use the shifted Mellin variable:
\begin{align}\label{eq:MellinShift}
n&=N-1\,.
\end{align}  
This allows use to write the leading power logarithms of $\x$ as poles at $n=0$.
Subleading power terms then correspond to poles at $n=-1,-2,\ldots$. These logarithms of $\x$ are generated both in the coefficient functions $\bar H_{p}^{(\kappa)}$ and the anomalous dimensions $\mb\gamma^s$. As remarked above in \secn{twistvsx}, direct coupling of photons to collinear partons is power suppressed in the small-$\x$ region. Hence, 
at leading non-trivial order in the $\x\rightarrow 0$ limit\footnote{More precisely, the  $n\rightarrow 0$ limit. However, for convenience, we will call this expansion the small-$\x$ expansion or small-$\x$ limit even in Mellin space.}, we have the expansions for the DIS electromagnetic structure functions:
\begin{align}\label{eq:LLxpoles}
	\bar H_{p}^{(\kappa)}\Big(n,\frac{Q^2}{\mu^2},\alpha_s(\mu^2)\Big) &= \frac{\alpha_s(\mu^2)}{\pi}\Bigg(\sum_{\ell=1}^{\infty} h_{p,\ell}^{(\kappa)}\Big(\frac{Q^2}{\mu^2}\Big)\bigg(\frac{\alpha_s(\mu^2)}{\pi n}\bigg)^{\ell} + \ldots\Bigg)+ \ldots \,,\\
	\gamma_{gg}\big(\alpha_s(\mu^2),n\big)&=\sum_{\ell=1}^{\infty} \gamma_{gg,\ell-1}\Bigg(\frac{\alpha_s(\mu^2)}{\pi n}\Bigg)^{\ell}+\ldots \,,	\nn\\
	\gamma_{qg}\big(\alpha_s(\mu^2),n\big)&=\frac{\alpha_s(\mu^2)}{\pi}\sum_{\ell=0}^{\infty} \gamma_{qg,\ell}\Bigg(\frac{\alpha_s(\mu^2)}{\pi n}\Bigg)^{\ell}+\ldots\,,	\nn
\end{align} 
where $\gamma_{\kappa \kappa'}(\alpha_s,n)$ are elements of the anomalous dimension matrix $\mb \gamma^s(\as,n)$.
We have neglected terms that are either logarithmically or power-suppressed suppressed as $\x\rightarrow 0$. As we will see below, in Mellin space, the above expansion of $\bar H_{p}^{(\kappa)}$ in the small-$\x$ limit at leading power only begins at $\cO(\as^2)$ due to the presence of an intermediate soft sector, and we find the behavior $h_{p,\ell}^{(\kappa)} \sim x^{-1} \ln^{\ell -1}(x)$, after the inverse Mellin transform is taken. Here we have indicated the leading logarithmic (LL) series for the coefficient function and the anomalous dimensions $\gamma_{gg}$ and $\gamma_{qg}$. In all cases we have a single logarithmic series: each power of $\alpha_s$ brings with it at most one additional logarithm. For the coefficient function and glue-to-quark anomalous dimensions, we have adopted the convention of Ref.~\cite{Catani:1994sq} (where the coefficient function is formally considered to start at next-to-leading logarithmic order, describing the same terms with different terminology than what we use here). This is due to the fact at fixed order, the small-$\x$ singularity requires an emission of a gluon, but at leading order in the DIS fixed order calculation, we can only probe quark intermediate states. As discussed below in \secn{resummation}, other terms in the anomalous dimension matrix, $\gamma_{\kappa q}$ for $\kappa=q,g$, are straightforwardly related to the above ones at LL. As detailed in Ref.~\cite{Catani:1994sq}, these quantities are relevant for the resummation of the DIS process at the first non-trivial order that small-$\x$ logs appear.

\subsection{Resummation by Catani and Hautmann}
\label{sec:CH}
The resummation of the glue-to-glue anomalous dimension at LL level was worked out long ago, Ref.~\cite{Jaroszewicz:1980mq}, and the extension of the resummation of the anomalous dimensions at higher orders in the BFKL approximation has been investigated in Ref. \cite{Fadin:1998py}. Further, it has been argued such resummations of the anomalous dimensions are not perturbatively small, necessitating a resummation of the BFKL kernel itself,  see Refs.~\cite{Ciafaloni:1998iv,Salam:1998tj,Ciafaloni:1999yw,Ciafaloni:2003rd,Ciafaloni:2003kd} and \cite{Altarelli:1999vw,Altarelli:2001ji,Altarelli:2003hk,Altarelli:2005ni} and \cite{Thorne:2001nr}. Resummation of the coefficient function and the gluon-to-quark anomalous dimension in the $\overline{\text{MS}}$-scheme was worked out in Ref.~\cite{Catani:1994sq}, and further investigated in Refs. \cite{Ciafaloni:2005cg,Ciafaloni:2006yk}. In general, the resummation follows from demanding consistency between the BFKL resummation of the DIS cross-section, and the factorization of the DIS structure functions in terms of parton distribution functions.

Here, we briefly review the BFKL and DGLAP resummation achieved by Catani and Hautmann (CH) in \Ref{Catani:1994sq} at LL accuracy. 
This allows us to review concepts and establish notation which will be useful for later comparisons with the EFT based approach. Their leading logarithmic factorization yields the following formula for the $F_L$ structure function for the gluon channel:
\begin{align}\label{eq:FLgCH}
	\bar F_{L}^{(g)}	\bigg( n ,\frac{Q^2}{\mu^2}\bigg)	=	h_L\big(	\gamma_{gg}(\as, n )\big)R(\as, n )	\Big(\frac{Q^2}{\mu^2}\Big)^{\gamma_{gg}(\as, n )}	\bar\Gamma_{gg}\big(\as 	,	n ,\eps\big)
\end{align}
As above, the IR divergences at leading power in small-$\x$ are absorbed in the transition function $\bar \Gamma_{gg}$ such that the remaining quantities in the formula above are IR-finite. 
Here, the function $h_L(\gamma)$ describes the coupling of photon to the incoming proton and is defined in terms of an $\cO(\as)$ off-shell cross section $\hat \sigma_L$ describing interaction of photons with space-like incoming gluons, with a $p = L$ projection:
\begin{align}\label{eq:hLdef}
	h_L(\gamma)	=	\gamma \int_0^\infty \frac{\df \mb k_\perp^2}{\mb k_\perp^2}	\: \Big(\frac{\mb k_\perp^2}{Q^2}\Big)^{\gamma}	\: 	\hat \sigma^g_{L}\bigg(\frac{\mb k_\perp^2}{Q^2},	\as ,	\eps =	0	\bigg)
  \,.
\end{align}

The anomalous scaling of the structure function in \eq{FLgCH} is governed by $\gamma_{gg}$ in \eq{LLxpoles}. CH achieved the resummation of $(\as/n)^\ell$ terms through a separate calculation involving an IR-divergent \textit{gluon Green's function} $\cF_g^{(0)}(n,k_\perp,	\as , \mu ,\eps)$, which  at the lowest order equals $\delta^{(2-2\eps)}(k_\perp)$ and satisfies the BFKL equation in $d -2 = 2- 2\eps$ dimensions.  The $g$ subscript on $\cF_g^{(0)}$ indicates that this is the gluon channel relevant for $F_L^{(g)}$ structure function. The resummation of $F_L^{(q)}$ structure function for quark channel is achieved via the quark channel of gluon Green's function $\cF^{(0)}_q$ which is straightforwardly related to $\cF_g^{(0)}(n,k_\perp,	\as , \mu ,\eps)$ in terms of ratio of Casimirs $C_F/C_A$.
Here $\epsilon$ regulates IR divergences, and we will discuss this in more detail later on.
Accordingly, all the higher order terms in $\cF_g^{(0)}$ are generated by the $d-2$ dimensional BFKL equation. Furthermore, the resulting series in $\as/n$ is IR divergent and CH showed that these IR divergences in $\cF_g^{(0)}$ can also be factored completely into $\bar\Gamma_{gg}$:
\begin{align}\label{eq:GluonGreen}
	\cF_g^{(0)}	\big(n, \mb k_\perp , \as ,\mu, \eps\big)	=	\frac{\gamma_{gg}(n,\as)}{\pi \mb k_\perp^2}	\tilde R\big(n, \mb k_\perp,	\mu_F,\as; \mu ,\eps \big)	\bar \Gamma_{gg}\big(\as ,	 n ,\eps,	\big)	\,	.
\end{align}
Note that factorization of $1/\eps$ poles in the structure function $\bar F_L^{(g)}$ in $\bar \Gamma_{gg}$ is a consequence of twist factorization.  On the other hand, the fact that $\bar \Gamma_{gg}$ captures all the $1/\eps$ IR poles in the gluon Green's function $\cF_g^{(0)}$, and that the offshell cross section $h_L(\gamma)$ in \eq{hLdef}  is collinear finite for $\eps \ra 0$ (for $\gamma \neq 0$), are special properties of the leading small-$\x$ logarithms. 
In \eq{GluonGreen} $\mu_F$ is the factorization scale for IR singularities.
 The term $\tilde R$ is IR-finite and describes the scheme dependence of IR factorization, and its $\eps \ra 0$ limit gives  $R(\as, n)$ in \eq{FLgCH}  multiplied with the scaling  factor $\big(\frac{\mb k_\perp^2}{\mu^2}\big)^{\gamma_{gg}(\as, n )}$. The prefactor $\gamma_{gg} $ in \eq{GluonGreen} is required to reproduce the classical $1/n$ scaling of the Green's function, the structure function, and the coefficient function as shown in \eq{LLxpoles}. Further imposing consistency with DGLAP  and BFKL resummation of $\cF_g^{(0)}$ enabled CH to obtain a closed form for $\gamma_{gg}$ via the implicit equation,
\begin{align}\label{eq:gammagg}
	1 = \frac{\as C_A}{\pi n }	\chi\big(\gamma_{gg}( n,\as)\big)	\,	,\qquad \chi(\gamma)	=	2\psi(1)	-	\psi(\gamma)	-	\psi(1-\gamma)	\,	,
\end{align}
and simultaneously solve for $R (\as, n )$ in \eq{FLgCH}. 

For the case of the $F_2$ structure function,  CH showed that in the limit $n\ra 0$, the coefficient function satisfies the following relation at LL accuracy:
\begin{align}
	\gamma_{gg}(N,\as)	\bar H_{2}^{(g)}	\big(n, Q^2/\mu^2 = 1,\as\big)	+ 2 n_f \gamma_{qg}\big( \as , n\big)	=	h_{2}(\gamma )	R\big(n,\as\big)	 \,	,	
\end{align}
where $h_2(\gamma)$ is defined analogously to \eq{hLdef} as 
\begin{align}\label{eq:h2def}
	h_2(\gamma)	=	\gamma \int_0^\infty \frac{\df \mb k_\perp^2}{\mb k_\perp^2}	\: \Big(\frac{\mb k_\perp^2}{Q^2}\Big)^{\gamma}	\: \frac{\partial}{\partial \ln Q^2}	\hat \sigma_{2}^g	\bigg(\frac{\mb k_\perp^2}{Q^2},	\as ,	\eps =	0	\bigg)	\,	.
\end{align}
The key difference for the $p = 2$ case in the approach of CH is that here, unlike the $p = L$ projection in \eq{hLdef}, the $\cO(\as)$ offshell kernel $\hat \sigma_2^g$ contains a $1/\eps$ IR pole that is unrelated to the IR divergences associated with BFKL evolution seen in \eq{GluonGreen}. This pole is related to the $\gamma_{qg}$ anomalous dimension and consequently the derivative with respect to $\ln Q^2$ renders it IR finite.  
However, this implies that unlike the $\gamma_{gg}$, the resummation of small-$\x$ logarithms in the coefficient function $\bar H_{2}^{(g)}	$ and $\gamma_{qg}$ cannot be achieved via the gluon Green's function alone in \eq{GluonGreen}. To this end, CH defined a \textit{quark Green's function},
\begin{align}\label{eq:QuarkGreen}
	G^{(0)}_{qg}\big( n,\as , \eps\big)	=	\int \df^{d-2}	\mb k_\perp  \: \hat K_{qg}\bigg(\frac{\mb k_\perp ^2}{Q^2}	,	\as , \mu , \eps\bigg)	\cF^{(0)}_g\big( n, \mb k_\perp ,\as , \mu ,\eps\big)	\,,
\end{align}
where the off-shell kernel $\hat K_{qg}$ captures the $1/\eps$ pole in $\hat \sigma_{2}^g$. As a result, similar to \eq{GluonGreen}, this formulation allowed CH to absorb the IR singularities in the quark Green's function $G^{(0)}_{qg}$  into $\bar \Gamma_{gg}$ and $\bar \Gamma_{qg}$ transition functions. Consistency with DGLAP resummation then enables determination of $\gamma_{qg}$ anomalous dimension, although not in a closed form as in \eq{gammagg}.

%%%%%%%%%%%%%%%%%%%%%%%%%%%%%%%%%%%%%%%%%%%%%%%%%%%%%%%%%%%%%%%%%%%%%%%%%%%%%%%
\section{EFT Modes and power counting}
\label{sec:modes}
%%%%%%%%%%%%%%%%%%%%%%%%%%%%%%%%%%%%%%%%%%%%%%%%%%%%%%%%%%%%%%%%%%%%%%%%%%%%%%%%
We briefly review the setup of SCET with Glauber operators in \secn{scet}.
 In \secn{pc} we describe the necessary and sufficient conditions for constraining the DIS process to be in the forward scattering regime and argue the necessity for an intermediate soft sector at leading power in this regime. The key differences in the twist and small-$\x$ expansions are discussed in \secn{twistvsx}.

%%%%%%%%%%%%%%%%%%%%%%%%%%%%%%%%%%%%%%%%%%%%%%%%%%%%%%%%%%%%%%%%%%%%%%%%%%%%%%%%
\subsection{Review of SCET and Glauber Operators}
\label{sec:scet}

To perform resummation of small-$x_b$ logarithms we will exploit the tools of soft collinear effective theory (SCET) with Glauber operators. SCET is a low energy effective theory with energetic quarks and gluons that encodes the appropriate soft and collinear degrees of freedom to adequately describe QCD in the infrared. The ``infrared'' region is defined by the 
appropriate separation of scales for a given problem. In SCET the soft and collinear degrees of freedom are distinguished by their momentum scaling chosen in appropriate light-cone coordinates. 
After appropriate factorization steps, at leading power in SCET the soft and collinear fields only have self-interactions, except for soft-collinear interactions mediated by Wilson lines in hard scattering operators or interactions mediated by Glauber operators.
Following the coordinate decomposition in \eq{LCDef}, we define the soft and collinear fields to obey the scalings given by 
\begin{align}
	&\text{soft:} & &k_s^\mu  \sim \sqrt{s}(\lambda,\lambda,\lambda) \,, & 
	&\text{$n$-collinear:}& &p_n^\mu \sim \sqrt{s}(\lambda^2, 1, \lambda) \, ,&
	&\text{$\bn$-collinear:}& &p_n^\mu \sim \sqrt{s}(1, \lambda^2, \lambda) \, .&
\end{align}
Here $\lambda$ is a small power counting parameter that we will identify below as a combination of kinematic variables of DIS, and $s$ is the center-of-mass energy of the scattering process.\footnote{Traditionally in the SCET literature, the symbol $Q$ is understood as the large momentum of the problem. We forgo this, as in the small-$\x$ regime the $Q$ in DIS will set the small momentum scale relative to $\sqrt{s}$. The large momentum scale of the collinear sectors is then $\sqrt{s}$.} These scalings imply that the ``virtuality'' or the off-shellness of the soft and collinear particles is constrained to $k_s^2 \sim p_{n,\bn}^2 \sim s\lambda^2$. From the top-down perspective of matching QCD onto SCET, any degrees of freedom with $p^2 \gg s\lambda^2$ are integrated out. As can be seen by simply adding the momentum components, direct interactions between soft and collinear fields (that is, at a single local vertex in a feynman diagram) render some propagators off-shell at leading power. For example, summing $k_s^\mu + p_n^\mu \sim \sqrt{s}(\lambda, 1 ,\lambda^2)$, we have the off-shellness $(k_s+p_n)^2\sim s\lambda \gg s\lambda^2$.

As we discuss shortly, the small-$\x$ regime corresponds to near forward scattering of the electron and proton in the center-of-mass frame with an intermediate soft sector. Here the hierarchy of scales results from wide separation in the rapidities (or light-cone momentum along the $n$ and $\bar{n}$ directions) of the three sectors involved, instead of large momentum transfer as in a hard scattering. The Glauber Lagrangian of \Ref{Rothstein:2016bsq} provides the necessary formalism for describing this regime. The interactions between soft and collinear modes are described by non-local operators given by
\begin{align}
	{\cal L}_{G}^{\rm II (0)} = e^{-\im  x\cdot {\cal P}} \sum_{ij} \sum_{n_i n_j } {\cal O}_{n_isn_j}^{ij} (x ) + e^{-\im  x\cdot {\cal P}}   \sum_{ij} \sum_{n_i} {\cal O}_{n_i s}^{ij} (x ) 
\end{align}
where the sum extends over the collinear and soft sectors widely separated in rapidity in directions $n_i$ and $n_j$ for partons $i,j = q,g$. The label momentum operator ${\cal P}^\mu$ appears as a result of having performed multipole expansion, and picks out large ${\cal O}(1)$ and ${\cal O}(\lambda)$ momentum components. We briefly review these operators in \app{GlauberAction}.

%%%%%%%%%%%%%%%%%%%%%%%%%%%%%%%%%%%%%%%%%%%%%%%%%%%%%%%%%%%%%%%%%%%%%%%%%%%%%%%%
\subsection{Power Counting for $\x\rightarrow 0$}
\label{sec:pc}

For our EFT analysis we will always assume that $\sqrt{s}$ is the largest momentum scale, and $\Lambda_{\rm QCD}$ is the smallest. 
The intrinsic bound quarks and gluons in the proton have transverse momentum $\sim \Lambda_{\rm QCD}$, so in the adopted $e^-$-proton center of mass frame the proton constituents have momenta with a collinear scaling
\begin{align} \label{eq:pcproton}
	p_c^\mu \sim \sqrt{s} \Big(\frac{\Lambda_{\rm QCD}^2}{s},1,
     \frac{\Lambda_{\rm QCD}}{\sqrt{s}}\Big) 
  \,. 
\end{align}
Here $p_c^2 \sim \Lambda_{\rm QCD}^2 \ll s$.  
DIS is often analyzed using the twist expansion,
where all operators with the lowest twist contribute equally to the OPE.
The twist expansion is an expansion in the power counting parameter
\begin{align} \label{eq:twistexpn}
  \lambda' \sim \frac{\Lambda_{\rm QCD}}{Q} \,.
\end{align}
For the usual $\lambda'$ expansion, no scaling is assigned to $s$ with respect to $Q^2$ or to $\x$. The result at leading twist is then neatly packaged into parton distribution functions yielding the formulas given in \eq{structure_f_factorization}. For the derivation of these factorization formula from SCET, see~\cite{Bauer:2002nz}. To start we will not enforce an expansion in $\lambda' \ll 1$, but will make use of \eq{twistexpn} in our discussion below.

To setup expansions for the small-$x$ region we will need to determine the scaling for $\x$ and $Q^2/s$. Defining $\lambda \ll 1$ as a small expansion parameter we let
\begin{align}
  \x\sim \lambda \,.
\end{align}
To determine the physical implications of possible choices for the scaling of $Q^2/s$, it is best to start from an assigned power counting for $P_X^2/s$.  The two scalings we consider are
\begin{align}\label{eq:scaling}
 & \text{\underline{i) Forward scattering:}} \nn\\*
&\qquad\quad
  \frac{P_X^2}{s} = \frac{(q+P)^2}{s} =\frac{Q^2}{s} \frac{(1- \x)}{\x} \sim   \lambda \, 
	\quad\Longrightarrow\quad
	\x \, y = \frac{Q^2}{s}\sim \lambda^2\,, \qquad
	y \sim \lambda \,. \quad
\end{align}  
\begin{align}\label{eq:scaling2}
  & \text{\underline{ii) Hard $P_X^2$:}} \nn\\*
&\qquad\quad
	\frac{P_X^2}{s} = \frac{(q+P)^2}{s} =\frac{Q^2}{s} \frac{(1- \x)}{\x} \sim   \lambda^0 
	\quad\Longrightarrow\quad
	\x \, y = \frac{Q^2}{s}\sim \lambda\,, \qquad
	y \sim \lambda^0 \,. \quad
\end{align} 
It is tempting to say that both scalings lead to forward scattering, since $Q^2$ is analogous to the standard Mandelstam variable $-t$, and in both cases $Q^2 \ll s$. 
This would suffice for \emph{elastic} forward scattering, where specifying that the exchanged particle carries small invariant mass is sufficient to specify the forward scattering regime. However, in DIS the process is inelastic, and  small $Q^2$ is a necessary but not sufficient condition.  Since $P_X^2 \sim s$ in the power counting ii), the final state invariant mass is in the hard momentum region, and there is necessarily always a hard scattering interaction at leading power.  In contrast, with the power counting in i) we have $P_X^2 \ll s$ and there will be no hard scatting interaction at leading power. This makes option i) the true scaling for the forward scattering limit. 

In principle small-$x$ resummation could be explored by adopting either scaling i) or ii), we wish to exploit knowledge of the EFT description of forward scattering described in \secn{scet}, and hence will adopt the scaling i) in \eq{scaling} for our analysis. With the choice of i) we can rewrite the scaling for the proton constituents in 
\eq{pcproton} as
\begin{align}\label{eq:col_pc}
		p_c=&\big ( p_c^+ , p_c^- ,p_{c\perp}\big )\sim \sqrt{s}\Big((\lambda\lambda^{\prime})^2,\,
		1,\, \lambda\lambda'\Big)\,.
\end{align}
Consistency requires $y\sim \lambda$ for this case, but the DIS structure functions $F_{2,L}(x,Q^2)$ are independent of $y$, and the full dependence on this variable can be entirely captured in the cross section prefactors. Using \eqs{qmu}{qperpSq} we can then determine that
\begin{align}\label{eq:photon_pc}
	q=&\big (q^+ , q^- ,q_{\perp}\big) \sim \sqrt{s} \Big(\lambda,\lambda^2,\lambda\Big)\,.
\end{align}  
This has the scaling of an $\bn$-$s$ Glauber mode in the language of Ref.~\cite{Rothstein:2016bsq}.\footnote{In contrast with the choice ii) we would have had $q\sim \sqrt{s}(1,\lambda,\sqrt{\lambda})$ for the virtual photon.
Essentially, the photon now has small invariant mass since it is collinear to the electron, not due to being in the forward scattering region.
Here $p_c \sim \sqrt{s}(\lambda \lambda^{\prime\,2}, 1, \sqrt{\lambda}\lambda')$ and $P_X^2 \sim (p_c+q)^2\sim s$ is in the hard region.}
Although we do not expand the leptonic tensor when deriving the factorization result, it is important to know the scaling of $q$ as it feeds momentum into $W^{\mu\nu}$ and hence effects the description of the modes used in its factorization. 

Crucially, taking $\lambda\ll 1$ this forces the leading contributions in the small-$x$ limit to have a soft-intermediate sector, as the Glauber photon cannot interact directly with the collinear sector of the proton without throwing it off-shell. This is true irrespective of whether or not take $\lambda'\sim 1$ or $\lambda'\ll 1$ in \eq{col_pc}. We thus require modes that can simultaneously couple to the $q$ in \eq{photon_pc} which has $q^+\sim q_\perp\sim \sqrt{s}\lambda$, and be on-shell so that they can cross a final state cut. This requires that we introduce soft modes with momentum scaling
\begin{align}\label{eq:soft_pc}
	p_s=&\big ( p_s^+ , p_s^- ,p_{s\perp}\big ) \sim \sqrt{s}\Big(\lambda,\lambda,\lambda\Big)\,.
\end{align}
Note that this is also consistent with \eq{scaling} since any radiation in $X$ is now either soft or collinear and 
\begin{align}
	P_X^2\sim(p_c+p_s)^2\sim p_c^- p_s^+ \sim s\lambda\,.
\end{align}  
Since the determination of the relevant modes for the $\lambda$ expansion does not depend on the $\lambda'$ expansion, our strategy will be to first consider the small-$x$ forward scattering expansion with $\lambda\ll 1$, treating $\lambda'\sim 1$. For this first stage, instead of the scaling in \eq{col_pc}, we therefore have collinear modes with the scaling
\begin{align} 
  p_n =&\big ( p_n^+ , p_n^- ,p_{n\perp}\big )\sim \sqrt{s}\Big(\lambda^2,\,
    1,\, \lambda\Big) \,.
\end{align}
These collinear modes with $p_{n\perp}\sim \sqrt{s}\lambda \sim Q$ are needed for the small-$x$ resummation. Only later will we then consider the twist expansion with $\lambda'\ll 1$. For this second $\lambda'$ expansion we could in principle include additional EFT modes with invariant masses $p^2\sim \Lambda_{\rm QCD}^2$.  This would not only include the proton constituent modes with momenta $p_c$ in \eq{pcproton}, but also additional modes, since as we will see later on, the soft modes in \eq{soft_pc} also contribute to the IR divergences present in the PDFs. This is due to the fact that regardless of how we power count $\lambda'$, integration over the intermediate Glauber exchanges between the soft and collinear sectors range over all scales below $Q^2$, simultaneously forcing the soft and collinear sectors to probe the infra-red. Rather than introducing these $p^2\sim \Lambda_{\rm QCD}^2$ modes explicitly, we will instead simply match the results from our first EFT onto the standard leading twist factorization formulae in \eq{structure_f_factorization}.

Finally, with the scaling in \eq{scaling} both projections in \eq{F2FLExp} for $F_L(x_b,Q^2)$  and $F_2(x_b,Q^2)$ give leading power contributions, in particular $(n\cdot q)^2/\qsq = Q^2/(s \x^2(1-y))\sim \lambda^0$ and $Q^2/(n\cdot q)^2= x_b^2 s/Q^2\sim \lambda^0$. Therefore the $x^2 n^\mu n^\nu$ term can be dropped relative to $\bn^\mu\bn^\nu$ for $F_2$, and we can define the leading power projectors: 
\begin{align}\label{eq:F2FLproj}
	\mathcal{P}_{p}^{\mu\nu}=\begin{cases} \frac{(n\cdot q)^2}{2\qsq} \, \bn_\mu \bn_\nu ,
    \qquad p = 2\\
		\frac{2Q^2}{(n\cdot q)^2}\, n_\mu n_\nu ,\qquad p = L
	\end{cases}
	\,.
\end{align}  

\subsection{The Twist expansion and Power Suppressed Contributions as $\x\rightarrow 0$}
\label{sec:twistvsx}

As emphasized, for the small-$\x$ forward scattering limit, we are free to make no assumption about the relative scaling of $\Lambda_{\rm QCD}^2$ over $Q^2$ (the expansion in $\lambda'$). One might be tempted to therefore conclude that the effective theory we construct from the $\lambda$ expansion with \eq{scaling} is valid \emph{eoipso} at all twists. However, the true situation is more intricate: many of the diagrams in the full theory that contribute at leading twist are power-suppressed in the small-$\x$ effective theory, and thus dropped in our leading power analysis of the small-$\x$ limit.\footnote{We also expect the converse to be true: the effective theory for small-$\x$ factorization will generate diagrams that are suppressed in the leading twist limit but are leading power at small-$\x$. As we will see below, whether or not these contributions are seen in a calculation depends on the regularization procedure for IR divergences, as well as the boundary conditions for BFKL evolution.} This will play a role in our analysis of the leading logarithmic resummation of the space-like DGLAP anomalous dimension $\mb\gamma^s$ in \secn{resummation}. We will see that in order to fully reconstruct the leading twist anomalous dimensions, we must match our small-$\x$ power expansion to the twist expansion, thereby including terms that are necessary for the renormalization of the operators in the twist expansion, but are themselves power-suppressed at small-$\x$.

\begin{figure}[t!]
	\begin{center}
		\includegraphics[width=0.5\textwidth]{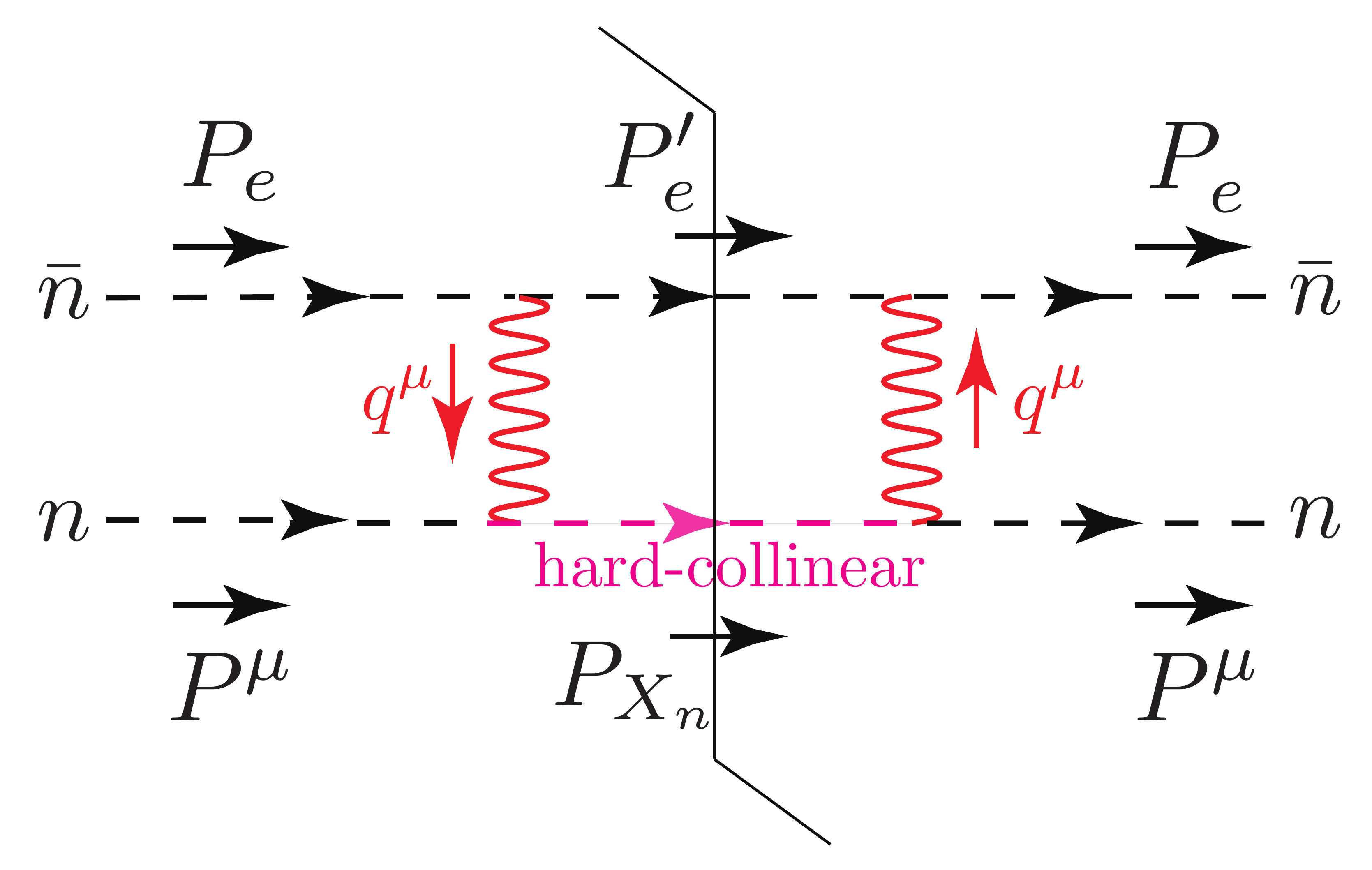} 
	\end{center}
	\vspace{-0.6cm}
	\caption{
		A direct interaction that does not involve a soft function at scale $q_\perp \sim Q$ and contributes only to $F_2$ and not to $F_L$. The intermediate collinear line will be off-shell (hard-collinear). }\label{fig:direct_term}
\end{figure}

In \fig{direct_term} we show an example of a tree-level diagram that contributes at leading twist and also at leading order in $\as$ for the coefficient function $H_2^{(\kappa)}$ in \eq{structure_f_factorization}. However, it does not appear in the same manner as the leading power terms in our small-$\x$ forward scattering expansion shown in \eq{LLxpoles} with the scaling defined by \eq{scaling}. This is because when a Glauber photon, with scaling as in \eq{photon_pc}, couples directly to the collinear sector, we end up with off-shell hard-collinear propagator with $p_{hc}^2\sim s \lambda$. This off-shell line is integrated out into a matching coefficient proportional to $H(z)\propto \delta(1-z)$, and hence yields a result that does not have a simple power counting in a strict $z\sim x$ expansion.  This can also be seen in moment space where $\int dz\, z^n \delta(1-z) = 1$ different from \eq{MellinPole}, where terms suppressed by $\sim \lambda^k$ relative to leading power appear as poles at $n = k$ in the Mellin space. This can be contrasted with the soft-sector mediated interactions, which perturbatively generate terms with the scaling $\sim (\ln x)^k/x$. 

%%%%%%%%%%%%%%%%%%%%%%%%%%%%%%%%%%%%%%%%%%%%%%%%%%%%%%%%%%%%%%%%%%%%%%%%%%%%%%%% 
\section{Factorization}
\label{sec:factorization}
%%%%%%%%%%%%%%%%%%%%%%%%%%%%%%%%%%%%%%%%%%%%%%%%%%%%%%%%%%%%%%%%%%%%%%%%%%%%%%%%

The factorization is derived in two steps. First we must formally match the full QCD electromagnetic current $J^{\mu}(x)$ onto the current operator composed from soft fields. This is trivial to do, since the soft current operator in SCET has the exact same form as for full QCD, and the multipole expansion can be simply implemented by dropping the $\bn \cdot q \sim \lambda^2$ photon momentum component, which is subleading, cf.~\eq{photon_pc}.

Next we must expand to the first non-trivial order in the Glauber lagrangian which mediates interactions between the soft and the proton-collinear sector. As discussed in \app{GlauberAction}, the Glauber soft-collinear action can be written in the form:
\begin{align}\label{eq:Glauber_operator_positions_space}
    S_G^{ns}&=8\pi \alpha_s% \big(\iota \mu^2\big)^{\frac{4-d}{2}}  
    \sum_{i,j,A}\int \df^d y \int \df^dx \int \frac{\df^d q'}{(2\pi)^d}  \frac{e^{\im (x-y)\cdot q'}}{\qpsq} {\cal O}_{n}^{iA}(x){\cal O}_{s}^{j_n A}(y)\,.
\end{align}
At this point we have already introduced dimensional regularization. 
This is necessary since even at tree level where the soft and collinear functions are both IR finite, the convolution in the factorization formula in \eq{fact} generates logarithmic IR divergences (we elaborate further on this below).
As will be seen from explicit calculation in \secn{CollFunc}, the collinear function is infrared-divergent beyond tree level, with a subset of the infrared divergences tied to the rapidity logarithms generated by the BFKL equation, a fact noted in Ref.~\cite{Catani:1994sq}. Additional IR divergences are generated when integrating over the momentum of the Glauber gluons in the convolution between soft and collinear sectors. Conversely, from rapidity renormalization group consistency, this  implies that the soft sector will also contain IR divergences at higher orders. Thus we are forced to introduce an IR regulator throughout the factorization. Since dimensional regularization affords the most convenient scheme for handling IR divergences, we will work throughout in $d=4-2\epsilon$ dimensions. Additionally, by choosing to work in dimensional regularization we necessarily set all the power divergences, and hence contributions from higher twist operators to zero. 

In \eq{Glauber_operator_positions_space}, we have written the operators completely in position space, and have combined all label sums with residual integrals, so labels are ``continuous.'' This enables us to simplify the derivation of the collinear function. Note that by momentum conservation and the requirement to respect the soft and collinear sector's power counting, both $\bar{n}\cdot q'$ and $q_\perp'$ are to be interpreted as $\cO(\sqrt{s}\lambda)$ quantities in the power counting, while $n\cdot q'$ is $\cO(\sqrt{s}\lambda^2)$. Then the matrix element in \eq{WmunuDef} to the lowest non-trivial order in the Glauber action for unpolarized target, as pictured in \fig{DISFact}, is given by
\begin{align}\label{eq:getting_started}
    W^{\mu\nu}&= \frac{1}{4\pi}\!\sum_{i_{L},j_{L},A_{L}}\int\df^d\!y_{L}\int  d ^d x_{L}\int \frac{\df^4q_L}{(2\pi)^d} \: \frac{e^{\im (x_{L}\cdot q_L-y_{L}\cdot q_L)}}{\vec q_{L\perp}^{\,2}} \\
    &\quad \times
    \sum_{i_{R},j_{R},A_{R}}  
	\int \df^d y_{R} \int \df^d x_{R}\!\int\! \frac{\df^dq_R}{(2\pi)^d}\:  \frac{e^{-\im (x_{R}\cdot q_R- y_{R}\cdot q_R)}}{\vec q_{R\perp}^{\,2}} \nonumber\\
    &\quad\times (8\pi \alpha_s)^2\int \df^d z\: e^{ \im z\cdot q}
    \langle P|\bar{T}\{J^{\mu}(z) {\cal O}_{n\,A}^{i_{L}}(x_{L}){\cal O}_{s\,A}^{j_{Ln}}(y_{L}) \}T\{J^{\nu}(0) {\cal O}_{n\,B}^{i_{R}}(x_{R}){\cal O}_{s\,B}^{j_{Rn}}(y_{R})\}|P\rangle	\nn\\
    &\quad +\ldots\nn \,.
\end{align}
We have neglected both power corrections in $\x$ and also higher order insertions of the Glauber action, which suffices for the NLL analysis that we perform here. We now begin our initial factorization by rewriting \eq{getting_started} as
\begin{align}\label{eq:almost_done}
    W^{\mu\nu}\!&=\frac{1}{4\pi}\int\!\frac{\df^d q_L}{(2\pi)^{d}}\frac{1}{\vec q_{L\perp}^{\,2}}  \int \frac{\df^d q_R}{(2\pi)^{d}}\frac{1}{\vec q_{R\perp}^{\,2}}\: 
   \mathcal{S}^{\mu\nu}_{AB}(q,q_L,q_R)\, 
   \mathcal{C}_{AB}(P,q_L,q_R) +\ldots   \,,
\end{align} 
where we have defined
\begin{align}\label{eq:cScCdef}
	    \mathcal{S}^{\mu\nu}_{A B}(q,q_L,q_R)&\equiv (8\pi \alpha_s)^2\sum_{i,j}\int \df^d z\: e^{\im q\cdot z} \int  \df^d y_L \df^d y_R\: e^{\im (- q_L\cdot y_L+ q_R\cdot y_R)} 
	\\
	&\quad \times \langle      0|\bar{T}\{J^{\mu}(z)\mathcal{O}_{s\,A}^{i_n}(y_L)\}T\{J^{\nu}(0)\mathcal{O}_{s\,B}^{j_n}(y_R)\}|0\rangle\,, \nn \\
	\mathcal{C}_{A B}(P,q_L,q_R)&\equiv \sum_{i,j}\int \df^d x_L\int \df^d x_R \: e^{\im x_L\cdot q_L- \im x_R\cdot q_R}\langle P|\mathcal{O}_{n\, A}^{i}(x_L)\mathcal{O}_{n\,B}^{j}(x_R)|P\rangle\,.    \nn
\end{align}
Note that we wrote a generic injection of all the components of $q_L$ and $q_R$ into the collinear function to make notation compact, and kept track of the color index associated to each Glauber operator with the momentum injected. Now we can use translation invariance to eliminate the $q_R$ dependence in the collinear function. We write $\mathcal{O}_{n}^{j\,A}(x_R)=e^{\im \mathbb{P}\cdot x_R}\mathcal{O}_{n}^{j\,A}(0)e^{-\im \mathbb{P}\cdot x_R}$, then we shift $x_L\rightarrow x+x_R$ to get
\begin{align}\label{eq:collinear_function_next_to_next_to_final}
  \mathcal{C}_{AB}(P,q_L,q_R)&=(2\pi)^{d}\vec q_{L\perp}^{\,2}\delta^{(d)}\Big(q_L-q_R\Big)\frac{\delta^{A B}}{N_c^2-1} {\cal C}(P,q_L)\,,\\
  {\cal C}(P,q')&\equiv\frac{1}{ \qpsq}\sum_{i,j,A}\int \df^d x\:  e^{\im x\cdot q'} \langle P|\mathcal{O}_{n}^{i\,A}(x)\mathcal{O}_{n}^{j\,A}(0)|P\rangle\,.    
\end{align} 
Note that even though there appears to be a power suppressed contribution of Glauber gluon momentum to the large component of collinear momentum $P$, we could not expand the function $\mathcal{C}_{AB}(P,q_L,q_R)$ in $q_L$ and $q_R$ in \eq{cScCdef}, because as seen in \eq{collinear_function_next_to_next_to_final} the $q_{L,R}$ dependence appears as an overall delta function $\delta^{(d)}(q_L- q_R)$ which is in fact homogeneous in the power-counting. This reflects the fact that the Glauber Langrangian of Ref.~\cite{Rothstein:2016bsq} cannot be expanded further in the multipole expansion. Expanding too soon in intermediate steps would lead to an uncontrolled volume factor $\delta^d(0)$.   Plugging in \eq{collinear_function_next_to_next_to_final} into \eq{almost_done}, we have
  \begin{align}\label{eq:factor_final_next_to_soft}    
    W^{\mu\nu}&=\frac{1}{4\pi}\int\!\frac{\df^d q'}{(2\pi)^{d}}{\cal S}^{\mu\nu}(q,q') \,{\cal C}(P,q')+ \ldots \,,\\
    {\cal S}^{\mu\nu}(q,q')&=\frac{(8\pi \alpha_s)^2}{(N_c^2-1)}\sum_{i,j,A}\!\int\! \df^d  z \df^dy_L \df^dy_R \: \frac{e^{\im z\cdot q-\im  q'\cdot (y_L-y_R)}}{ \qpsq}	\nn\\
    &\quad \times 
    \langle 0|\bar{T}\{J^{\mu}(z)\mathcal{O}_s^{i_n\,A}(y_L)\}T\{J^{\nu}(0)\mathcal{O}_s^{j_n\,A}(y_R)\}|0\rangle\,. \nn
  \end{align}  
Having extracted the momentum conservation $\delta$ function from the collinear sector which was homogeneous in the power counting, we now can expand the momentum convolution between the soft and the collinear sectors. The momentum $q'$ has the Glauber gluon scaling:
\begin{figure}[t]
    \begin{center}
        \includegraphics[width=0.6\textwidth]{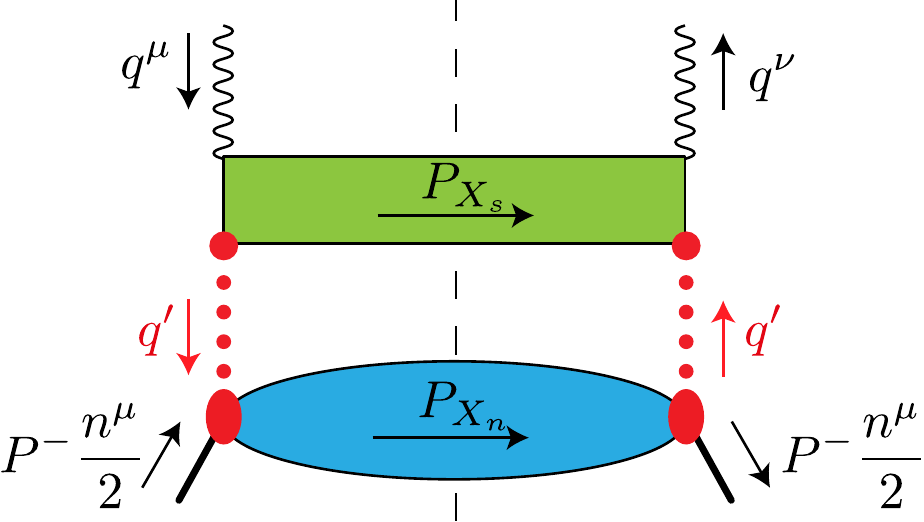}\hspace{10pt}
        \includegraphics[width=0.3\textwidth]{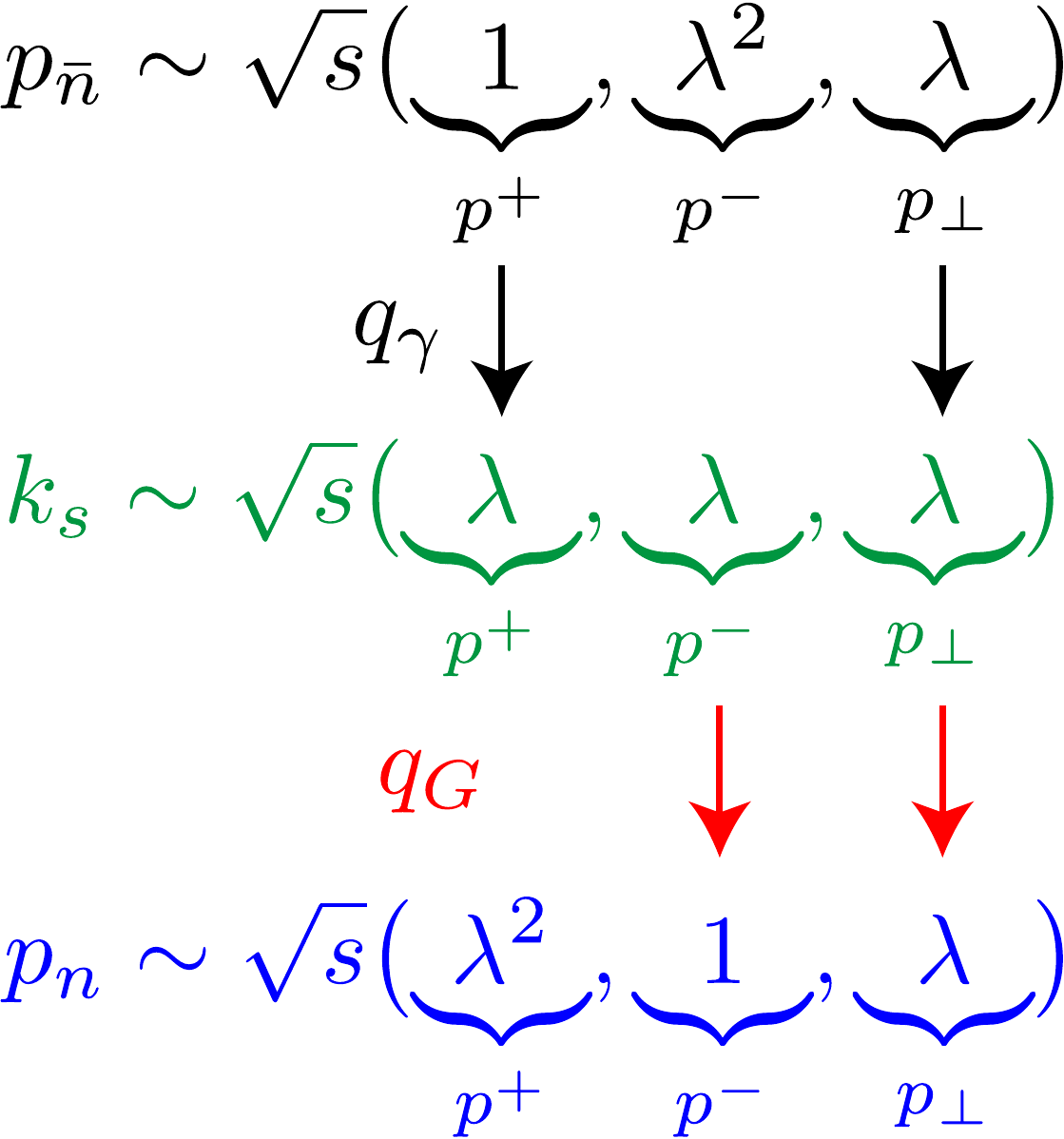}
    \end{center}
    \vspace{-0.6cm}
    \caption{ Factorization at lowest order in the Glauber exchange. }
    \label{fig:DISFact}
\end{figure}
\begin{align}
q' = \big(q^{\prime+}, q^{\prime-}, q_{\perp}'\big) \sim \sqrt{s}\big(\lambda^2, \lambda, \lambda \big) \,.
\end{align}  
In contrast with \eq{photon_pc}, the momentum $p$ has the scaling of $n$-$s$ Glauber exchange. Since in the soft function the $q^{\prime +}$ component is subleading, and within the collinear function the $q^{\prime-}$ component is subleading, we may set these to zero in these functions. Then applying the projectors from \eq{F2FLproj}, we achieve the following factorization for the structure functions:
\begin{align}\label{eq:WmunuFinal}
        \frac{1}{\x}F_p(\x,Q^2)&= \int\!\df^{d-2} q_\perp' \: S_p\Big(\frac{\nu n\cdot q}{\qsq},q_{\perp},q_\perp',\eps \Big)C\Big(\frac{\nu}{\bn\cdot P},q_\perp',\eps \Big)+\ldots \,, 
      \end{align}
where
\begin{align} 
\label{eq:factor_final_collinear}  
C\Big(\frac{\nu}{\bn\cdot P},q_\perp',\eps \Big)&\equiv \frac{1}{\pi \nu} \sum_{i,j,A} \int \frac{\df n\cdot q' }{2\pi} \: \int \df^dx \:  
\frac{e^{ \im \frac{\bn \cdot x}{2} n\cdot q' + \im x_\perp\cdot q_\perp'}}{ \qpsq} \langle P|\mathcal{O}_{n}^{i\,A}(x)\mathcal{O}_{n}^{j\,A}(0)|P\rangle_\nu 
\,, \\
\label{eq:factor_final_soft}
    S_p\Big(\frac{\nu n\cdot q}{\qsq},q_{\perp},q_\perp',\eps \Big)&\equiv \frac{(2\pi \iota \mu^2)^{4-d}\big(8\pi \alpha_s(\mu^2)\big)^2\nu}{16\pi^2 \,(N_c^2-1)} \sum_{i,j,A} \int\frac{\df \bn\cdot q'}{4\pi}\int \df^d z \:e^{\im z\cdot q} \int \df^d y_L \df^d y_R 	
     \nn   \\
    &\quad\times \frac{e^{ - \im \frac{\bar{n}\cdot q'}{2}(n\cdot y_L-n\cdot y_R) - \im q_\perp'\cdot (y_{L\perp}-y_{R\perp})}}{ \qpsq}	\nn\\
    &\quad \times
    \mathcal{P}_{p\, \alpha\beta }\langle 0|\bar{T}\{J^\alpha(z)\mathcal{O}_s^{i_n\,A}(y_L)\}T\{J^\beta(0)\mathcal{O}_s^{j_n\,A}(y_R)\}|0\rangle_\nu\,.
\end{align}  
Here the $\nu$ indicates the presence of a rapidity regulator, and $S_p\Big(\frac{\nu n\cdot q}{\qsq},q_{\perp},q_\perp',\eps\Big)$ and $C\Big(\frac{\nu}{\bn\cdot P},q_\perp',\eps\Big)$ are rapidity renormalized soft and collinear functions. We indicate the extra steps involved in this renormalization by the subscript $\nu$ on the matrix elements. Due to boost invariance (RPI-III), the dependence on $n\cdot q$ and $\bn\cdot P$ appears only due to rapidity divergences, which in the regulated functions results in the combinations shown in the first arguments of $S_p$ and $C$.  We have explicitly included the $(\iota \mu^{2})^{(4-d)}$ that comes from the integration over the two Glauber lines into the soft function, as we associate the corresponding $\alpha_s$ factors to the soft function. The coupling is expressed in the $\overline {\rm MS}$ scheme with $\iota$ given by
\begin{align}\label{eq:MSbar}
	\iota  \equiv  \frac{ e^{\gamma_E}}{4\pi} \, .
\end{align}
Next, the $\nu$ in the prefactors in the collinear and soft functions in \eqs{factor_final_collinear}{factor_final_soft} are convenient to reproduce the  expected classical small-$x$ scaling and we will see this below explicitly in \secn{collFuncLL}. With these definitions, the soft and collinear functions have the (classical) mass dimensions:
\begin{align}\label{eq:dim}
    \big[ S^{\mu\nu} \big] = 4-d \, , \qquad \big[C\big] = -2  \, .
\end{align}
At this point, we can derive another form of the factorization theorem from how the collinear function must scale in dimensional regularization. The motivation for this form is from the fact that the integration on the product of the soft and collinear contributions in \eq{WmunuFinal} is itself infra-red divergent. Thus we cannot set $d=4$, and naively perform the integration against the rapidity renormalized soft and collinear functions. We can however make these infra-red divergences explicit in the soft function, by writing out a more explicit form of the collinear function as a loop expansion in dimensional regularization. Since rapidity integrals and transverse momentum integrals factorize from each other, the all-orders form of the collinear function in dimensional regularization  can be written as:\footnote{This follows because  dimensional regularization only changes the transverse dimensions, thus the scaling in $\mu^{2\epsilon}$ must follow the scaling in the injected transverse momentum, up to corrections induced by the running coupling constant. This is true for a correctly zero-bin subtracted~\cite{Manohar:2006nz} collinear function in \SCETb in a reasonable rapidity regularization scheme.}
\begin{align}\label{eq:collinear_function_expand}
C\Big(\frac{\nu}{\bn\cdot P},q_\perp',\eps \Big)&=\sum_{\ell=0}^{\infty}\frac{1}{\qpsq}C^{(\ell)}\Big(\frac{\nu}{\bn\cdot P},\alpha_s(\mu^2),\epsilon\Big)\Big(\frac{\mu^2}{\qpsq}\Big)^{\ell\epsilon}\,.
\end{align}  
Note that while we have assumed that we have inserted the counterterms to appropriately renormalize the coupling, we have not expanded in $\epsilon$. Thus \eq{collinear_function_expand} is UV finite but IR divergent, and the coefficients $C^{(\ell)}$ do not have a specific scaling form in $\alpha_s$ due to its renormalization. However, when we solve the BFKL equation, we will be able to derive the leading logarithmic expression for $C^{(\ell)}\big(\frac{\nu}{\bn\cdot P},\alpha_s(\mu^2),\epsilon\big)$ and find that $C^{(\ell)} \sim \as^\ell$. 

\begin{figure}[t!]
	\centering
	\includegraphics[width=0.5\textwidth]{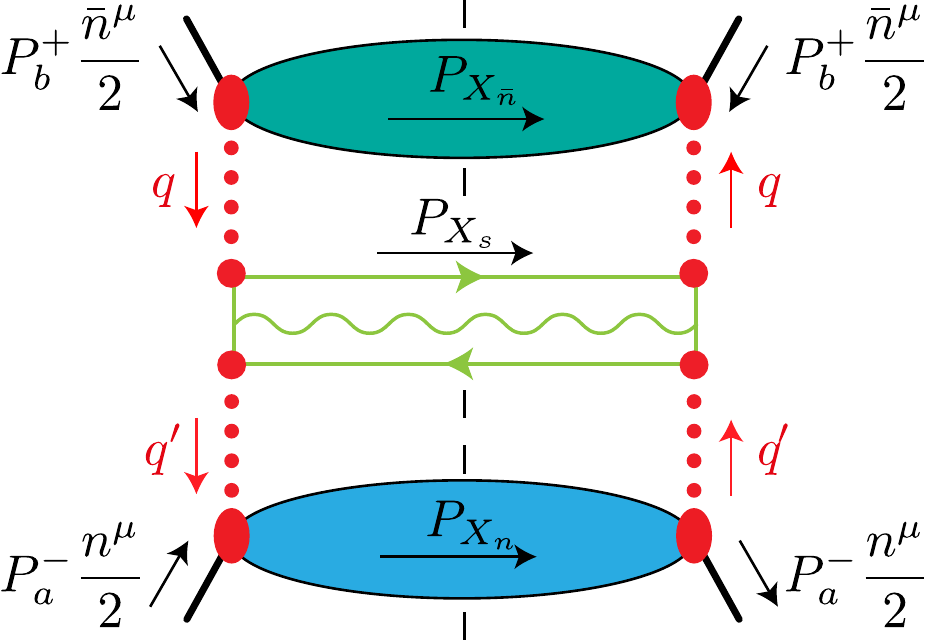}
	\vspace{-6pt}
	\caption{Forward scattering regime in Drell-Yan. The collinear region in the directions $n$ and $\bn$ is described by the same collinear function $C$ studied in this work. The green cut propagators result in  soft quarks and a soft photon in the final state.}
	\label{fig:DY}
\end{figure}

We are now in a position to derive specific forms of the factorization useful for the resummation of the form factors. We can define a ``Mellin-transformed'' soft function, with respect to the transverse momentum, to trade the integration for a summation:
\begin{align}\label{eq:Gamma_form_factorization}
  \frac{1}{\x}F_p(\x,Q^2)&=  \sum_{\ell=0}^{\infty} \Big(\frac{\qsq}{\mu^2}\Big)^{-(\ell + 2)\epsilon}C^{(\ell)}\Big(\frac{\nu}{\bn\cdot P},\alpha_s(\mu^2),\epsilon\Big)\tilde{S}_p\Big(\frac{\nu n\cdot q}{\qsq},-\ell\epsilon,\alpha_s(\mu^2),\eps\Big)+\ldots \,,
\end{align}
where we have defined 
\begin{align}
  \Big(\frac{\qsq}{\mu^2}\Big)^{\gamma-2\epsilon}\tilde{S}_p\Big(\frac{\nu n\cdot q}{\qsq},\gamma,\alpha_s(\mu^2),\eps\Big)&\equiv\int \frac{d^{2-2\epsilon}q_\perp^{\prime}}{\qpsq}\Big(\frac{\qpsq}{\mu^2}\Big)^{\gamma}S_p\Big(\frac{\nu n\cdot q}{\qsq},q_{\perp},q_\perp',\eps \Big)\,.\label{eq:Gamma_form_soft}
\end{align}
The scaling in transverse momentum, corresponding to the prefactor we have extracted on the left hand side of \eq{Gamma_form_soft}, follows from 
dimensional analysis for the $\gamma$-transform of the definition in \eq{factor_final_soft}. From \eq{dim} we see that $\tilde S_p$ is dimensionless.

Finally, we note that the collinear function defined in \eq{factor_final_collinear} is process independent and universal. Thus, the same function also appears in other forward scattering processes. The process dependence is described through the soft function. As an illustration, we consider the Drell-Yan process. For the hadrons to undergo forward scattering the produced photon must be soft. A leading order diagram is shown in \fig{DY}. The intermediate soft sector is coupled to the collinear sectors via Glauber exchanges. The $n$ and $\bn$-collinear sectors involve precisely the same collinear function we defined above.

\section{Fixed Order Calculations}
\label{sec:fixedorder}

Having derived the factorization formula in \eq{WmunuFinal} and the operator definitions of the soft and collinear functions in \eqs{factor_final_collinear}{factor_final_soft} we turn to their fixed order computations. In \secn{softLO} we compute the DIS soft function $S^{\mu\nu}$ at tree level in momentum and Mellin space, and in \secn{CollFunc} we compute the collinear function to next-to-leading order (NLO). These calculations will provide the basic ingredients we require for performing LL small-$\x$ resummation of the coefficient function and the anomalous dimensions in \secn{resummation}. While the LO result of the collinear function already suffices for LL small-$\x$ resummation,  our NLO computation will provide us with a cross check of the form of the BFKL equation governing the rapidity logarithms. We will make use of the rapidity regulator of Refs.~\cite{Chiu:2011qc,Chiu:2012ir}.

%%%%%%%%%%%%%%%%%%%%%%%%%%%%%%%%%%%%%%%%%%%%%%%%%%%%%%%%%%%%%%%%%%%%%%%%%%%%%%%%
\subsection{Soft Function at LO}
%%%%%%%%%%%%%%%%%%%%%%%%%%%%%%%%%%%%%%%%%%%%%%%%%%%%%%%%%%%%%%%%%%%%%%%%%%%%%%%%
\label{sec:softLO}
We now illustrate the structure of the momentum flow within the soft function by calculating its lowest order contribution. We first rewrite the expression in \eq{factor_final_soft} by inserting complete sets of soft states, shifting the argument of $J^\mu(z)$ and performing the $z$ integral:
\begin{align}
	&S_p \Big(\frac{\nu n\cdot q}{\qsq}, q_\perp, q_\perp',\eps \Big)  = \nu\frac{\alpha_s^2 (2\pi\iota \mu^2)^{2\eps}}{(N_c^2 - 1) \qpsq} \sum_{i,j,A} \sum_{X_s}
	 \int \frac{\df \bn \cdot q'}{\pi}
	\int \df^d y_L \df^d y_R \: 
	\delta^{(d)} \big(P_{X_s}^\mu - q^\mu + q^{\prime \mu} \big)\mathcal{P}_{p\,\rho\sigma}	\nn
	 \\
	& 
	\times
	e^{-\im \frac{\bn \cdot q'}{2} n\cdot (y_L - y_R) - \im q_\perp' \cdot (y_L - y_R)_\perp} \langle 0 | \overline {\rm T} \big \{ J^\alpha(0) \: {\cal O}_s^{i_n\,A}(y_L)\big \} | X_s \rangle \langle X_s | {\rm T} \big \{J^\beta(0)\: {\cal O}_s^{j_n\,A} (y_R)\big \} | 0 \rangle  \, .
\end{align}
Here $q^{\prime\mu} $ is the momentum transferred to the $n$-collinear sector and $q^\mu$, the incoming photon momentum, and we recall the projector $\mathcal{P}_{p\,\rho\sigma}$ to the correct structure function ($p=2,L$) in \eq{F2FLproj}. Setting the subleading components of the incoming photon and Glauber momenta to zero, we have
\begin{align}
	q^{\prime \mu}  = \frac{\bn \cdot q'}{2} n^\mu + q_\perp^{\prime\mu} \, ,
	\qquad     q^{\mu}=\frac{n\cdot q}{2}\bar{n}^{\mu}+q_{\perp}^{\mu} \, .
\end{align}
\begin{figure}[t!]
	\centering
	\includegraphics[width=0.8\textwidth]{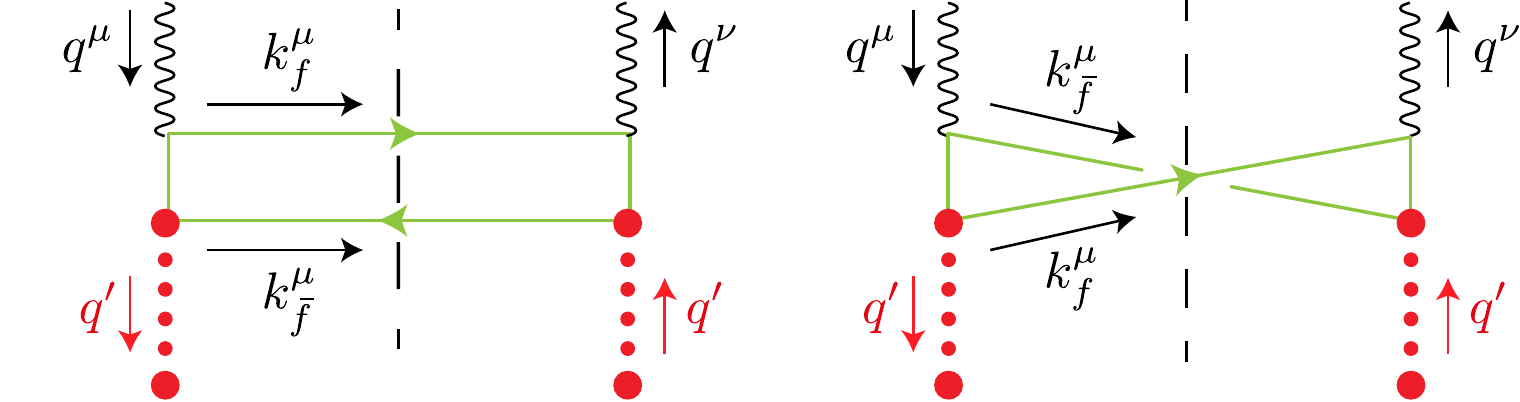}
	\caption{Box and crossed-box diagrams for the tree level soft function. $q^\mu$ is the incoming photon momentum and $q^{\prime \mu}$ the Glauber momentum exchanged with the $n$-collinear sector (not shown). Two more diagrams are obtained by interchanging quark and anti-quark lines and $k_f$, $k_{\bar f}$ momenta.}
	\label{fig:soft_tree}
\end{figure}
Thus, the final state soft particles carry the remaining momentum $P^\mu_{X_s} = q^\mu - q^{\prime\mu}$. 

At the lowest order, the final state is $X_s = q\bar q$. The two LO diagrams are shown in \fig{soft_tree}. We let $k_{f}$ and $k_{\bar{f}}$ be the momenta of the quark and anti-quark crossing the cut, and $k_L$ and $k_R$ be the virtual momenta of the quark propagator on either side.  Then we have,
\begin{align}\label{eq:sf_start}
    &S_p(n\cdot q,q_{\perp},q_\perp',\eps )=\frac{\alpha_s^2 P^-(2\pi\iota \mu^2)^{2\eps}}{\qpsq}\mathcal{P}_{p\, \alpha\beta}\!\int\!\frac{\df \bn \cdot q'}{\pi}\!\int[\df^d k_f]_{+}\!\int[\df^d k_{\bar{f}}]_{+}\!\int\frac{\df^d k_L}{(2\pi)^d}\int\frac{\df^d k_R}{(2\pi)^d}\\
    &\quad\times\Big((2\pi)^{3d}\delta^{(d)}(q-k_f+k_L)\delta^{(d)}(-q'-k_{\bar{f}}-k_L)\delta^{(d)}(q'+k_{\bar{f}}-k_R) \text{box}^{\alpha\beta} (k_f,k_{\bar{f}},k_L,k_R)\nonumber\\
    &\quad\quad+(2\pi)^{3d}\delta^{(d)}(q-k_{\bar{f}}+k_L)\delta^{(d)}(-q'-k_{f}-k_L)\delta^{(d)}(q'+k_{f}-k_R) \text{box}^{\alpha\beta} (k_{\bar{f}},k_{f},k_L,k_R)\nonumber\\
    &\quad\quad-2(2\pi)^{3d}\delta^{(d)}(q-k_f+k_L)\delta^{(d)}(-q'-k_{\bar{f}}-k_L)\delta^{(d)}(q'+k_f-k_R) \text{c-box}^{\alpha\beta}(k_f,k_{\bar{f}},k_L,k_R)\Big) \, ,\nonumber
  \end{align}
where
\begin{align}
        [\df^d k]_{+}&\equiv\frac{\df^d k}{(2\pi)^d}2\pi\delta(k^2)\Theta(k^0) \, .
\end{align}
The delta functions result from the explicit vertices in the time-ordered products in the \eq{factor_final_soft}. We sum over the two ways the fermion lines can circulate, such that  the box and cross-box diagrams given by
\begin{align}\label{eq:box}
\text{box}^{\alpha \beta} (k_f,k_{\bar{f}},k_L,k_R)&=  n_f T_F \frac{\text{tr}[\slashed{k}_f\gamma^{\beta}\slashed{k}_R\slashed{n}\slashed{k}_{\bar{f}}\slashed{n}\slashed{k}_L\gamma^{\alpha}]}{(k_L^2+i\epsilon)(k_R^2-i\epsilon)}\,,\\
\text{cbox}^{\alpha \beta} (k_f,k_{\bar{f}},k_L,k_R)&= n_f T_F \label{eq:cbox}
\frac{\text{tr}[\slashed{k}_f\slashed{n}\slashed{k}_R\gamma^{\beta}\slashed{k}_{\bar{f}}\slashed{n}\slashed{k}_L\gamma^{\alpha}]}{(k_L^2+i\epsilon)(k_R^2-i\epsilon)}\,.
\end{align}  
After integrating over the momenta of the left and right propagators we get:
\begin{align}\label{eq:sf_II}
    &S_p\Big(\frac{\nu n\cdot q}{\qsq},q_{\perp},q_\perp',\eps \Big)= \frac{\alpha_s^2 \nu(2\pi\iota \mu^2)^{2\eps}}{\qpsq}\mathcal{P}_{p\,\alpha\beta}\int\frac{\df \bn \cdot q'}{\pi}\int[\df^d k_f]_{+}\int[\df^d k_{\bar{f}}]_{+}\delta^{(d)}(q-q'-k_f-k_{\bar{f}})	\nn\\
    &\qquad \times\Big(\text{box}^{\alpha \beta} (k_f,k_{\bar{f}},k_{f}-q,q'+k_{\bar{f}})+(\bar{f}\leftrightarrow f)-2\, \text{cbox}^{\alpha \beta}(k_f,k_{\bar{f}},k_{f}-q,q'+k_{f})\Big) \, .	
  \end{align}
Next, the phase space integral in \eq{sf_II} can be simplified as
\begin{align}\label{eq:PSsoft}
	&\int\frac{\df \bn \cdot q'}{\pi}\int[\df^d k_f]_{+}\int[\df^d k_{\bar{f}}]_{+} (2\pi)^d\delta^{(d)}(q-q'-k_f-k_{\bar{f}}) \\ 
	&\qquad = \frac{1}{n\cdot q} \int_0^1 \frac{\df w }{(2\pi)w (1-w)} \int \frac{\df^{d-2}\ell_\perp}{(2\pi)^{d-2}} \, , \nn 
\end{align}
where $\ell^\mu$ is momentum of one of the outgoing fermion, and we have defined $w = n\cdot \ell /n\cdot q $. 

Further, the computation of the $\gamma$-transformed dimensionless soft function  defined in \eq{Gamma_form_soft} makes the final result even simpler.
Our explicit results for $\tilde{S}_2$ and $\tilde{S}_L$ are:
\begin{align}
  \tilde S_{2}\Big(\frac{\nu n\cdot q}{\qsq},\gamma,\as(\mu^2),\eps \Big)&=
  \frac{1}{2}\frac{(2\pi \iota)^{2\eps}}{4^{1-2\epsilon}}
  \as^2n_fT_F\Big(\frac{\nu n\cdot q}{\qsq}\Big)\frac{\pi^2 \csc \big(\pi(\gamma-2\epsilon)\big)\csc\big(\pi(\gamma-\epsilon)\big)}{\Gamma(2-\epsilon)\Gamma(\frac{5}{2}-\gamma+\epsilon)\Gamma(\frac{3}{2}+\gamma-2\epsilon)} \nn\\
  &\quad\times\big(2-10\epsilon+4\epsilon^2+8\epsilon^3+\gamma(3+5\epsilon-12\epsilon^2)+\gamma^2(-3+4\epsilon)\big)\,,\nonumber\\
    \tilde S_{L}\Big(\frac{\nu n\cdot q}{\qsq},\gamma,\as(\mu^2),\eps\Big )&= -\frac{(2\pi \iota)^{2\eps}}{4^{1-2\epsilon}}
    \as^2n_fT_F\Big(\frac{\nu n\cdot q}{\qsq}\Big)	\nn\\
    &\quad \times \frac{\pi^2(\gamma-2\epsilon)(\gamma-2\eps - 1)\csc\big(\pi(\gamma-2\epsilon)\big)\csc\big(\pi(\gamma-\epsilon)\big)}{\Gamma(1-\epsilon)\Gamma(\frac{5}{2}-\gamma+\epsilon)\Gamma(\frac{3}{2}+\gamma-2\epsilon)}\,.
\end{align}
Once we have calculated the resummed collinear function, we will plug these expressions into \eq{Gamma_form_factorization} in \secn{collFuncLL} to find the LL small-$\x$ resummed structure functions. 

We pause to note that at finite photon and Glauber momenta, and likewise for finite $\gamma$, the above tree-level results are IR finite. However, the integration over transverse momentum of the exchanged Glauber ranges down to zero in the convolution with the collinear function in \eq{Gamma_form_factorization},  generating an IR divergence. In the $\gamma$-transformed soft finction, we can see the IR divergence explicitly by expanding at $\gamma = -\ell \eps\rightarrow 0$ following \eq{Gamma_form_factorization}, such that:
\begin{align}
	\label{eq:S2Leps}
	\lim\limits_{\eps \ra 0}	\tilde S_{2}\Big(\frac{\nu n\cdot q}{\qsq},-\ell \eps ,\as (\mu^2), \eps\Big)	&=\frac{2\as^2 n_f T_F}{3\pi} \frac{1}{(\ell + 1)(\ell + 2)}\bigg(\frac{1}{\eps^2}	+ \frac{2}{\eps} + \cO(\eps^0)\bigg)	\,	,	\\
	\lim\limits_{\eps \ra 0}	\tilde S_{L}\Big(\frac{\nu n\cdot q}{\qsq},-\ell \eps ,\as(\mu^2), \eps\Big)	&= \frac{2\as^2 n_f T_F}{3\pi}\frac{1}{(\ell + 1)}\bigg(-\frac{1}{\eps}	+ \cO(\eps^0)\bigg)	\,	.	\nn
\end{align}
This property of the soft function implies that not only it captures the process dependence, it also contributes to the PDF despite being a vacuum matrix element.

These results can be compared with the IR finite offshell cross sections $h_{2,L}(\gamma)$ of \Ref{Catani:1994sq}
introduced above in \eqs{hLdef}{h2def}. 
To obtain analogous terms from our setup
we keep $\gamma$ finite and set $\epsilon=0$, finding
\begin{align}\label{eq:S2Lh2L}
		\tilde S_{2}\Big(\frac{\nu n\cdot q}{\qsq},\gamma,\as ,\eps = 0\Big) &=   \Big(\frac{\nu n\cdot q}{\qsq}\Big)\alpha_s \frac{h_2(\gamma)}{\gamma^2} 
		\,, \\
		\tilde S_{L}\Big(\frac{\nu n\cdot q}{\qsq},\gamma,\as ,\eps = 0\Big) &=   \Big(\frac{\nu n\cdot q}{\qsq}\Big)\alpha_s\frac{h_L(\gamma)}{\gamma}  \nn \, .
\end{align}
As discussed in \secn{CH}, the offshell cross sections $\hat \sigma^g_L$ and $\hat \sigma^g_2$ that define $h_{L,2}(\gamma)$ in \eqs{hLdef}{h2def} exhibit different IR divergences. Since in the approach of \Ref{Catani:1994sq}, the IR divergences are captured separately by the auxiliary quark and gluon Green's functions given in \eqs{GluonGreen}{QuarkGreen},  
 the additional factors of $\gamma$ in the denominator in \eq{S2Lh2L} render the $h_{2,L}(\gamma)$ functions in the numerators IR finite. This difference in the relative powers of $\gamma$ for $p = 2$ and $L$ can also be directly compared to the difference in the $\eps \ra 0$ behavior of these two functions in \eq{S2Leps}. In contrast to \Ref{Catani:1994sq},  the full $\eps$ dependence of $\tilde S_2$ and $\tilde S_L$ is important for us to perform small-$\x$ resummation of anomalous dimensions and coefficient functions, and it reduces the number of independent ingredients that must be calculated to obtain final results.

Finally, we noted above that the offshell cross sections $\hat \sigma^g_L$ and $\hat \sigma^g_2$ are $\cO(\as)$. Since the leading power behavior in small-$\x$ limit is seen only starting from $\cO(\as^2)$, 
we will see below that $h_{2,L}(\gamma)$  are in fact related to power suppressed contributions
to the corresponding coefficient functions that are required for consistency of the small-$\x$ resummed structure function with twist factorization.
We continue the discussion of this in \secn{resummation} where we carry out the twist expansion.

%%%%%%%%%%%%%%%%%%%%%%%%%%%%%%%%%%%%%%%%%%%%%%%%%%%%%%%%%%%%%%%%%%%%%%%%%%%%%%%%
\subsection{Collinear Function at NLO}
%%%%%%%%%%%%%%%%%%%%%%%%%%%%%%%%%%%%%%%%%%%%%%%%%%%%%%%%%%%%%%%%%%%%%%%%%%%%%%%%
\label{sec:CollFunc}
We now discuss the computation of the collinear function at one-loop accuracy defined in \eq{factor_final_collinear}. In perturbation theory we replace the incoming proton states by partonic states $\kappa(p)$ carrying momentum $p$ where $\kappa = q,g$. For simplicity, in this section we replace the incoming Glauber momentum $q^{\prime\mu} \ra q^\mu$ below and continue to work in $d = 4-2\eps$ dimensions. Performing the $x^\mu$ integration and inserting complete sets of collinear states yields
\begin{align}\label{eq:CkappaDef}
	C_\kappa \Big(\frac{\nu}{\bn\cdot P}, \vec q_\perp^{\,2},\eps  \Big) &\equiv\frac{ 1}{\pi \nu} \frac{1}{2 N_\kappa} \frac{1}{\vec q_\perp^{\,2}} \sum_{i,j = q,g}  \int \frac{\df  n\cdot q}{2\pi} \\
    &\quad \times \sum_X (2\pi)^d \delta (P^- - \bn \cdot p_X) \delta (n\cdot q - n\cdot p_X) \delta^{(d-2)} (q_\perp - p_{X\perp})  \nn \\ 
	&\quad \times \langle \kappa(p) |  O_n^{Aj} (0)|X  \rangle \langle X | O_n^{Ai} (0) | \kappa(p)\rangle  \, .\nn
\end{align}
Written this way, we see that the incoming collinear quark or gluon with momentum $p^\mu = P^- n^\mu/2 = (0, P^- , 0_\perp)$ is struck by an off-shell gluon with momentum $q^\mu = (n\cdot q, 0, q_\perp)$.  The normalization $2 N_\kappa$ implies average over spins and colors of the initial partonic state with
\begin{align}
	N_q &= N_c = C_A 
	\,, \qquad N_g = N_c^2 - 1  = 2 C_A C_F
	\, .
\end{align}
Also note that the $n\cdot q$ momentum of the Glauber gluon is not resolved and integrated over. The $\sum_X$ denotes the phase space integral over the on-shell final-state particles in the state $X$ and is given by
\begin{align}
	\sum_X =\sum_{n=1}^{\infty}\int d{\rm PS}_n \equiv \sum_{n=1}^{\infty}\int  \Bigg[ \prod_{i = 1}^n \frac{\df^d  \ell_i }{(2\pi)^d} \: (2\pi) \Theta (\bar{n}\cdot\ell_i) \delta (\ell_i^2)\Bigg] \, .
\end{align}

At tree-level we only have a single collinear parton in the final state, and the phase space integral simplifies as
\begin{align}\label{eq:1PSDef}
	\int \frac{d  n\cdot q}{2\pi}  \: d	{\rm PS}_1  \:(2\pi)^d \delta^{(d)}\big(q + P - p_X\big) = \frac{1}{\bn \cdot P}  \, , \qquad n\cdot q = \frac{\vec q_\perp^{\,2}}{\bn \cdot P} \, .
\end{align}
The tree level matrix element simply yields $C_F C_A 2 (\bn \cdot P)^2$ such that
\begin{align}\label{eq:CnTree}
	C_q^{(0)} \Big(\frac{\nu}{\bn\cdot P}, \vec q^{\,2}_\perp\Big) &=  \frac{\bn\cdot P}{\nu}\frac{ C_F}{\pi \vec q_\perp^{\,2}} \,, 
	\qquad 
	C_g^{(0)} \Big(\frac{\nu}{\bn\cdot P}, \vec q_\perp^{\,2}\Big) =  \frac{\bn\cdot P}{\nu}\frac{ C_A}{\pi \vec q_\perp^{\,2}}  \,.
\end{align}

For the two particle phase space at ${\cal O}(\alpha_s)$, for the quark initial state we choose $k^\mu$ as the quark momentum and $\ell^\mu$ as the gluon momentum. For the gluon initial state, these momenta refer to the two outgoing gluons or two outgoing quarks.  We define the momentum fraction $z \equiv \ell^- /P^-$. Momentum conservation constrains $k^\mu$ as follows
\begin{align}
	n\cdot k &=  n\cdot q - n\cdot \ell  \, , \qquad
	\bn \cdot k = (1- z) P^- \,, \nn \qquad
	k_\perp = q_\perp - \ell_\perp \,.
\end{align}
The phase space integral for two outgoing partons simplifies as
\begin{align}
	\int \frac{d  n\cdot q}{2\pi}  \: d	{\rm PS}_2 \: (2\pi)^d \delta^d\big(q^\mu + P^\mu - p_X^\mu\big)  =
        \frac{1}{4\pi \bn\cdot P}\int_0^1 \frac{\df z}{z(1-z)} \int \frac{d^{2-2\eps}\ell_\perp}{(2\pi)^{2-2\eps}} \, ,
\end{align}
where the $n\cdot q$ momentum is constrained as
\begin{align}
	n\cdot q = \frac{(\vec q_\perp - \vec \ell_\perp)^2}{(1-z)\bn\cdot P} + \frac{\vec \ell_\perp^2}{z\bn\cdot P} \, .
\end{align}
%
%------------------
\subsubsection{Initial quark}
%------------------

\begin{figure}[t!]
			\centering
		\includegraphics[width=0.9\textwidth]{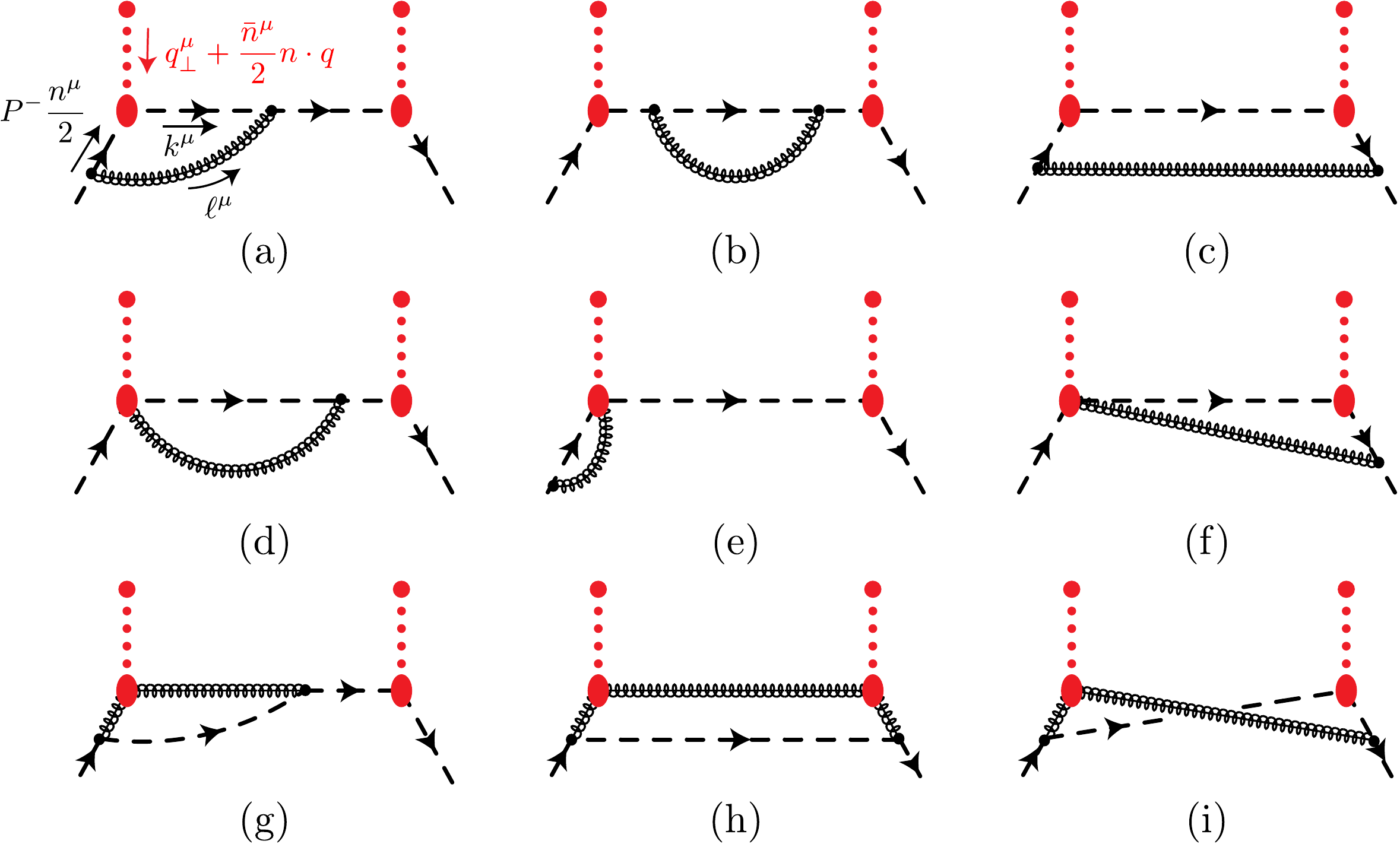}
	\vspace{-6pt}
	\caption{Graphs for incoming quark struck by an off-shell Glauber gluon. The two-parton final state cuts yield the real radiation graphs and the one-parton cuts the one-loop virtual corrections. The top row gives the $C_F^2$ terms. Graphs in bottom row involve $O_n^{gA}$ operator insertions. Graphs obtained by taking conjugate are not shown.}
			\label{fig:quarkCut}
\end{figure}
The real emission graphs for incoming quark are obtained by the cuts across the diagrams shown in \fig{quarkCut} that yield two partons in the final state. The collinear function is obtained by squaring the amplitude with the result being
\begin{align}
	C_q^{(1), \text{real}} \Big(\frac{\nu}{\bn\cdot P}, \vec q^{\,2}_\perp\Big) &= \frac{1}{\pi\nu} 	\int \frac{d  n\cdot q}{2\pi}  \: d	{\rm PS}_2 \: (2\pi)^d \delta^{(d)}\big(q + P - p_X\big) \:  {\cal C}_q^{(1), \text{real}} (p, q, \ell ,k )\, , 
\end{align}
where the integrand is given by
\begin{align}\label{eq:Cqreal}
	&{\cal C}_q^{(1), \text{real}}  = \bn\cdot P\frac{g^2  (\iota \mu^2)^{\eps}}{2} P_{gq}^\eta (z) \Bigg( \frac{C_F^2 \, z}{(k\cdot \ell)(\ell\cdot p)} +  \frac{C_F C_A}{2}  \bigg[ \frac{1}{k\cdot \ell}\bigg(\frac{-z}{l\cdot p} + \frac{(1-z)}{k\cdot p} \bigg)  +  \frac{1}{(\ell\cdot p)(k\cdot p)}\bigg]  
	 \Bigg) .
\end{align}
Here the pre-factor is the quark splitting function:
\begin{align}
	P_{gq}^\eta (z) =  \Big(\frac{\nu}{\bn\cdot P}\Big)^\eta \Big(\frac{1+(1-z)^2}{z^{1+\eta}}\Big) + {\cal O}(\eta) \, .
\end{align}
As we will show below, only the second term in the $C_FC_A$ piece in \eq{Cqreal} survives after including the virtual graph contributions, and leads to a rapidity divergent contribution. This term arises solely due the last, non-planar graph in \fig{quarkCut}(i) that involves one $O_{n}^{Aq}$ and one $O_{n}^{Ag}$ operator insertion.

Next, we express the dot products in terms of the transverse momenta
\begin{align}
	k \cdot \ell  = \frac{(\vec \ell_\perp - z\vec q_\perp )^2}{2z(1-z)} \, , \qquad 
	\ell \cdot p = \frac{\vec \ell_\perp^{\, 2}}{2z} \, , \qquad 
	k \cdot p = \frac{(\vec \ell_\perp - \vec q_\perp)^2}{2(1-z)} \, ,
\end{align}
such that
\begin{align}
	C_q^{(1), \text{real}}\Big(\frac{\nu}{\bn\cdot P}, \vec q^{\,2}_\perp\Big) &= 
	2 \frac{\alpha_s }{\pi}  (\iota \mu^2)^{\eps} \frac{\bn\cdot P}{\nu} \int_0^1 \df z \: 
 	P_{gq}^\eta (z) 
 		\int \frac{d^{2-2\eps}\ell_\perp}{(2\pi)^{2-2\eps}} \Bigg(
 	C_F^2 \frac{z^2}{\vec \ell_\perp^{\,2} (\vec \ell_\perp - z \vec q_\perp)^{\,2}}
 	\\
 	& \qquad
	+ \frac{C_F C_A}{2} \bigg[
	\frac{1}{(\vec \ell_\perp - z\vec q_\perp)^2} \bigg(-\frac{z^2}{\vec \ell_\perp^{\,2}} + \frac{(1-z)^2}{(\vec \ell_\perp -\vec q_\perp)^2}\bigg) + \bigg(\frac{1}{\vec \ell_\perp^{\,2}(\vec \ell_\perp -\vec q_\perp)^2}  \bigg)
	\bigg]
	\Bigg) \nn \, .
\end{align}
To simplify this further, we define the integral
\begin{align}\label{eq:Ieps}
	I_\eps \big[\vec r_\perp^{\,2}\big] &\equiv \Big(\frac{\mu^2 e^{\gamma_E}}{4\pi}\Big)^\eps
	\int \frac{d^{2-2\eps}\ell_\perp}{(2\pi)^{2-2\eps}} \frac{\vec r_\perp^{\,2}}{\vec \ell_\perp^{\,2} (\vec \ell_\perp + \vec r_\perp)^{\,2}}
	\nn \\
	&=  \Big(\frac{\vec r_\perp^{\,2}}{\mu^2}\Big)^{-\eps} \Gamma(-\eps) e^{\eps \gamma_E} \frac{\Gamma(1-\eps)\Gamma(1+\eps)}{(2\pi)\Gamma(1-2\eps)}  \nn \\
	&= \frac{1}{2\pi} \bigg[ \frac{1}{\eps} - \log\Big(\frac{\mu^2}{\vec q_\perp^{\,2}}\Big)\bigg] + {\cal O}(\eps)
	\, .
\end{align}
We will also make use of the relations
\begin{align}
	\Big(\frac{\mu^2 e^{\gamma_E}}{4\pi}\Big)^\eps
	\int \frac{d^{2-2\eps}\ell_\perp}{(2\pi)^{2-2\eps}} \frac{\vec r_\perp \cdot (\vec \ell_\perp + \vec r_\perp)}{\vec \ell_\perp^{\,2} (\vec \ell_\perp + \vec r_\perp)^{\,2}} &=
	\frac{1}{2} I_\eps[r_\perp^{\,2}]\, , \\
	\Big(\frac{\mu^2 e^{\gamma_E}}{4\pi}\Big)^\eps
	\int \frac{d^{2-2\eps}\ell_\perp}{(2\pi)^{2-2\eps}} \frac{\vec \ell_\perp \cdot (\vec \ell_\perp + \vec r_\perp)}{\vec \ell_\perp^{\,2} (\vec \ell_\perp + \vec r_\perp)^{\,2}} &=
	-	\frac{1}{2} I_\eps[r_\perp^{\,2}]\, , 
\end{align}
Thus, we have
\begin{align}\label{eq:CqReal}
		C_q^{(1), \text{real}}\Big(\frac{\nu}{\bn\cdot P}, \vec q^{\,2}_\perp\Big) &=  \frac{2 \alpha_s  }{\pi \vec q_\perp^{\,2}} \frac{\bn\cdot P}{\nu}  I_\eps\big[\vec q_\perp^{\,2}\big] \int_0^1 \df z \: 
		P_{gq}^\eta (z) \\
		&\qquad \times \bigg[ \Big(C_F^2 - \frac{C_F C_A}{2}\Big) z^{-2\eps} + \frac{C_F C_A}{2}\big( (1-z)^{-2\eps} + 1\big) \bigg] \, .
		\nn
\end{align}

\begin{figure}[t!]
	\centering
	\includegraphics[width=0.8\textwidth]{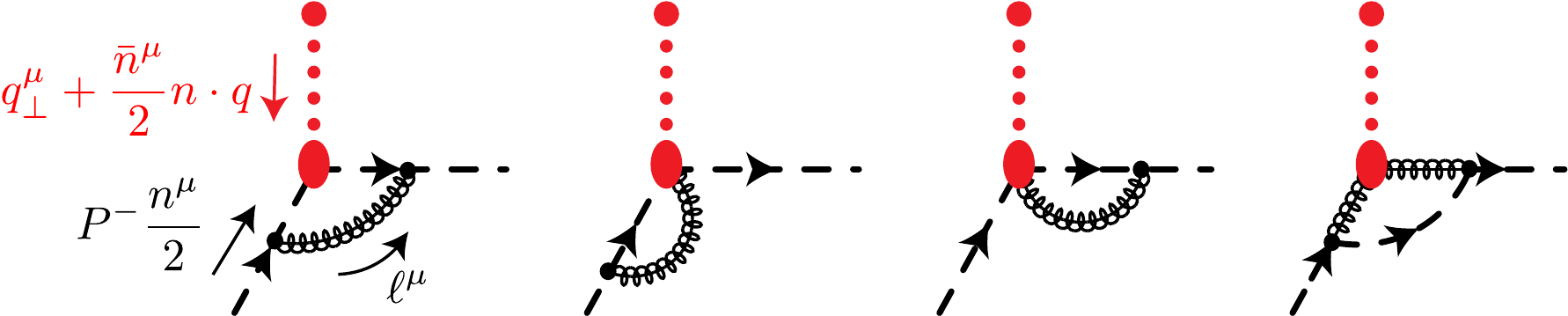}
	\vspace{-6pt}
	\caption{The virtual graphs for initial $n$-collinear quark struck by an off-shell Glauber gluon.}
	\label{fig:quarkV}
\end{figure}
We now turn to the virtual graphs obtained by taking one-particle final state cuts in \fig{quarkCut}, and shown in \fig{quarkV}. The vertex correction graph involving the $O_{n}^{Aq}$ (including the pre-factors in \eq{CkappaDef} and the one-particle phase space in \eq{1PSDef}) is given by
\begin{align}
	A_v =  \Big(C_F^2 - \frac{C_FC_A}{2}\Big) 2\im \frac{1}{\pi \vec q_\perp^{\,2}} g^2(\mu^2 \iota )^{\eps}\frac{\bn\cdot P}{\nu}\int \frac{\df^d  \ell}{(2\pi)^d} 
	\frac{(z-1)\vec q_\perp^{\, 2} + \vec \ell_\perp^{\, 2} - z\vec \ell_\perp \cdot \vec q_\perp}{[\ell^2 + \im 0 ][(p-\ell)^2 + \im 0] [(p-\ell +q)^2+ \im 0]}	\,	.
\end{align}
The denominators are given by
\begin{align}
	\ell^2 + \im 0 &= zP^- \Big[n\cdot \ell - \frac{\vec \ell_\perp^{\,2}}{z P^-} + \frac{\im 0}{z}\Big] \,,  \\
%	(\ell - q)^2 + \im 0 &= z P^- \Big[n\cdot \ell - n\cdot q - \frac{(\vec \ell_\perp - \vec q_\perp)^2}{z P^-} + \frac{\im 0}{z} \Big] \,, \nn \\
	(p-\ell)^2 + \im 0 &= (-1) (1-z)P^- \Big[n\cdot \ell + \frac{\vec \ell_\perp^{\,2}}{P^-(1-z)} - \frac{\im 0}{1-z}\Big] \, , \nn \\
	(p-\ell + q)^2 + \im 0 &= (-1) (1-z)P^- \Big[n\cdot \ell - n\cdot q + \frac{(\vec q_\perp - \vec \ell_\perp)^2}{P^-(1-z)} - \frac{\im 0}{1-z}\Big] \, .\nn
\end{align}
Hence, we have non-zero contribution for $0< z < 1$. Closing the contour below we get
\begin{align}
	A_v &=  \alpha_s  \Big(C_F^2 - \frac{C_FC_A}{2}\Big) 2\frac{ (\mu^2 \iota )^{\eps} }{\pi \vec q_\perp^{\,2}}\frac{\bn\cdot P}{\nu}\int_0^1 \df z  \int\frac{d^{d-2}\ell_\perp}{(2\pi)^{d-2}} \:  \frac{z \big[\vec \ell_\perp \cdot (\vec \ell_\perp - z\vec  q_\perp) - (1-z) \vec q_\perp^{\,2} \big]}{\vec \ell_\perp^{\,2}(\vec \ell_\perp - z\vec q_\perp)^2} \nn \\
	&=- \alpha_s   \Big(C_F^2 - \frac{C_FC_A}{2}\Big) \frac{1}{\pi \vec q_\perp^{\,2}}    I_\eps \big[\vec q_\perp^{\,2}\big]
	\int_0^1 \df z \:\frac{ \big[1 + (1-z)^2\big]}{z^{1+2\eps}}  \, .
\end{align}
The graphs in \fig{quarkV} involving Wilson line emission give scaleless contributions.  The last graph involving mixing with the $O_{n}^{Ag}$ operator gives
	\begin{small}
\begin{align}
	D_v &= \frac{ (\mu^2 \iota )^{\eps} }{2\pi \vec q_\perp^{\,2}} \frac{\bn\cdot P}{\nu}\int \frac{\df^d  \ell}{(2\pi)^d} \sum_s \overline u_n^s (p) \Big(\frac{\bnslash}{2}T^A\Big) (\psl+ \qslash)
\big(\im g \gamma_\nu T^C 
\big)\Big(\frac{-\im}{\ell^2}\Big)\Big(\frac{\im (\pslash - \ellslash + \qslash)}{(p-\ell + q)^2}\Big)
\frac{-\im}{(\ell - q)^2} (\im g \gamma_\mu T^E) u_n^s(p)\nn
\\
&\times (\mu^{2\eps} \iota^\eps  |\bn\cdot \ell|^{-\eta}\nu^\eta)
\im f^{EAC} \Big[\bn\cdot \ell \, g_\perp^{\mu\nu} - \bn^\mu (\ell_\perp-q_\perp)^\nu -\bn^\nu \ell_\perp^\mu + \frac{(\ell_\perp-q_\perp)\cdot \ell_\perp \,\bn^\nu \bn^\mu}{\bn\cdot \ell}\Big] \,.
\end{align}
	\end{small}
This time we close the contour above and find
\begin{align}
	D_v &= 2\alpha_s   \Big(\frac{C_F C_A}{2}\Big) \frac{ (\mu^2 \iota )^{\eps} }{\pi \vec q_\perp^{\,2}}  \frac{\bn\cdot P}{\nu}\int_0^1 \df z \:\frac{\big[1 + (1-z)^2\big] }{z^{1+\eta}}  \Big(\frac{\nu}{P^-}\Big)^\eta 
	\int\frac{d^{d-2}\ell_\perp}{(2\pi)^{d-2}} \: 
	\frac{(\vec \ell_\perp-\vec q_\perp)\cdot (\vec \ell_\perp  -z \vec q_\perp)}{(\vec \ell_\perp-\vec q_\perp)^2(\vec \ell_\perp  -z \vec q_\perp)^2} \nn \\
	&=-\alpha_s   \Big(\frac{C_F C_A}{2}\Big) \frac{1}{\pi \vec q_\perp^{\,2}}   I_\eps\big[\vec q_\perp^{\,2}\big] \Big(\frac{\nu}{\bn\cdot P}\Big)^{-1+\eta}\int_0^1 \df z \:\frac{\big[1 + (1-z)^2\big] }{z^{1+\eta}(1-z)^{2\eps}}    
\end{align}
The final result for the virtual contributions to the collinear function is twice the sum of $A_v$ and $D_v$. Thus, for an inclusive measurement we see that the $C_F^2$ pieces cancel completely. Combining with the result for real radiation graphs in \eq{CqReal} we have
\begin{align}\label{eq:quark1loop}
	 C_q^{(1)}\Big(\frac{\nu}{\bn\cdot P}, \vec q_\perp^{\,2},\eps \Big) &=  \alpha_s  C_F C_A \frac{1}{\pi \vec q_\perp^{\,2}}   \frac{\bn\cdot P}{\nu} I_\eps\big[\vec q_\perp^{\,2}\big] \int_0^1 \df z \: 
 	P_{gq}^\eta (z)  \nn \\
     &= \frac{\alpha_s C_A}{\pi} C_q^{(0)}\Big(\frac{\nu}{\bn\cdot P}, \vec q_\perp^{\,2}\Big) (-2\pi) I_\eps\big[\vec q_\perp^{\,2}\big]
     \bigg(\frac{1}{\eta} + \log\Big(\frac{\nu}{P^-}\Big) + \frac{3}{4}\bigg)  	\, .
\end{align}
%------------------
\subsubsection{Initial gluon}
%------------------
\begin{figure}[t!]
	\centering
	\includegraphics[width=0.8\textwidth]{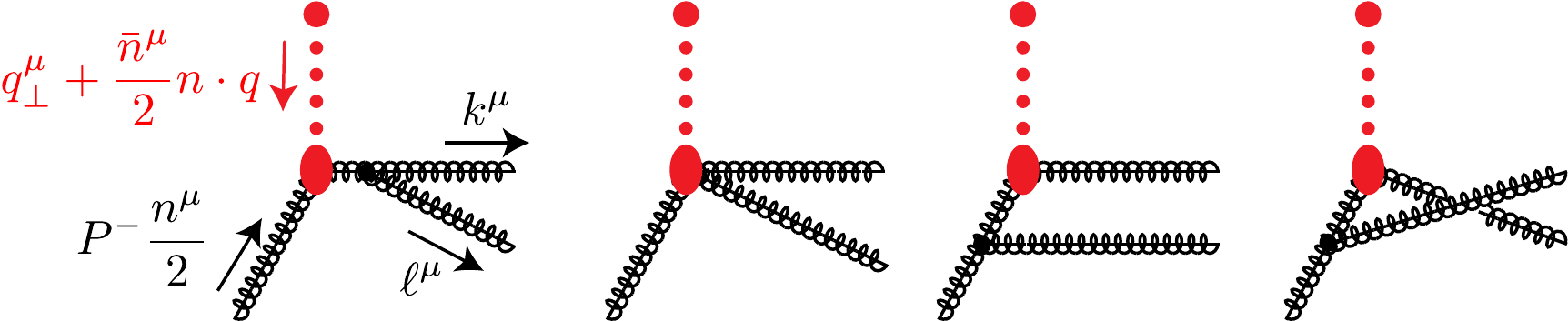}
        \includegraphics[width=0.8\textwidth]{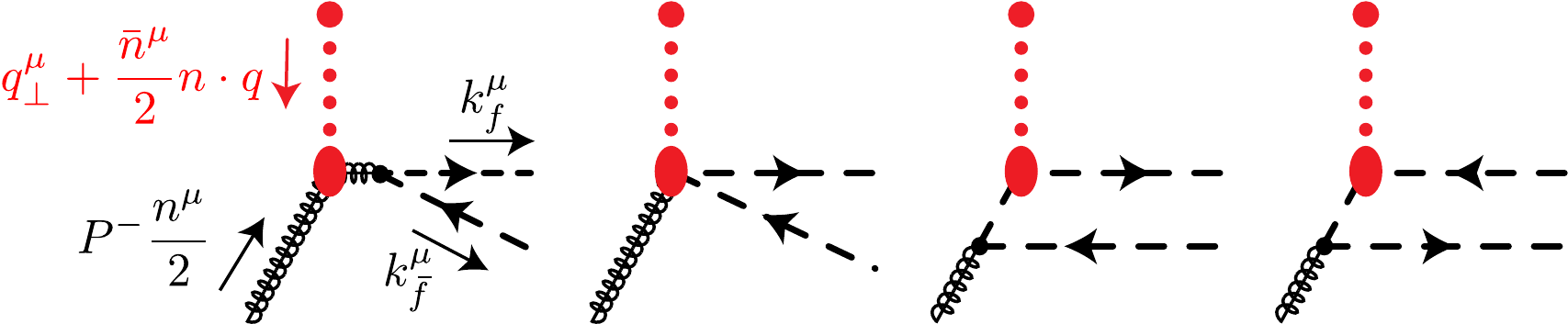}
    
	\vspace{-6pt}
	\caption{The real radiation graphs for initial $n$-collinear gluon struck by an off-shell Glauber gluon. The collinear function is obtained by squaring the amplitude and summing over the final state gluons colors and helicities.}
	\label{fig:gluonR}
\end{figure}
The real-emission graphs for an incoming gluon are shown in \fig{gluonR}. Squaring the amplitude and summing over outgoing helicities and colors we find
\begin{align}
	C_g^{(1), \text{real}}\Big(\frac{\nu}{\bn\cdot P}, \vec q^{\,2}_\perp\Big) &= \frac{1}{2\pi\nu }	\int \frac{d  n\cdot q}{2\pi}  \: d	{\rm PS}_2 \: (2\pi)^d \delta^d\big(q^\mu + P^\mu - p_X^\mu\big) \: \mu^{2\eps} {\cal C}_g^{(1), \text{real}} (p, q, \ell ,k )\, .
\end{align}
We have included a factor of $1/2$ for the symmetric final state. The integrand is given by
\begin{align}
	{\cal C}_g^{(1), \text{real}}  &= \bn\cdot P \frac{g^2\mu^{2\eps} C_A^2}{2}\Big( P_{gg}^{\eta}(z)+\frac{n_fT_R}{2C_A}P_{q\bar{q}}(z)\Big)
 \bigg(\frac{k\cdot \ell + z k\cdot p + (1-z) \ell \cdot p}{(\ell\cdot k)(k\cdot p)(\ell \cdot p)}\bigg) \, ,
\end{align}
with the splitting functions being\footnote{The diagrams containing the intermediate quark-antiquark states also have a contribution proportional to $C_F$. These are scaleless in both the real and virtual terms, and cancel anyways in the sum. For compactness of presentation, we drop such terms. }
\begin{align}
  P_{gg}^\eta(z) &= \frac{1 + z^4 + (1-z)^4}{2} \Big(\frac{\nu}{\bn\cdot P}\Big)^\eta \bigg(\frac{1}{z^{1+\eta}} + \frac{1}{(1-z)^{1+\eta}}\bigg) + {\cal O}(\eta) \, ,\\
  P_{q\bar{q}}(z) &= 1-2\frac{z(1-z)}{1-\epsilon}\,.
\end{align}
Including the phase space and integrating over the perp-momenta we find
\begin{align}
	C_g^{(1), \text{real}}\Big(\frac{\nu}{\bn\cdot P}, \vec q^{\,2}_\perp\Big) &= 
 \frac{	 \alpha_s C_A^2 }{\pi \vec q_\perp^{\,2}} \frac{\bn\cdot P}{\nu}   I_\eps\big[\vec q_\perp^{\,2}\big] \int_0^1\!\! \df z \: 
	\Big(P_{gg}^\eta (z)+\frac{n_fT_R}{2C_A}P_{q\bar{q}}(z)\Big) \big(1 + z^{-2\eps} + (1-z)^{-2\eps}\big)  .
\end{align}

\begin{figure}[t!]
	\centering
	\includegraphics[width=0.7\textwidth]{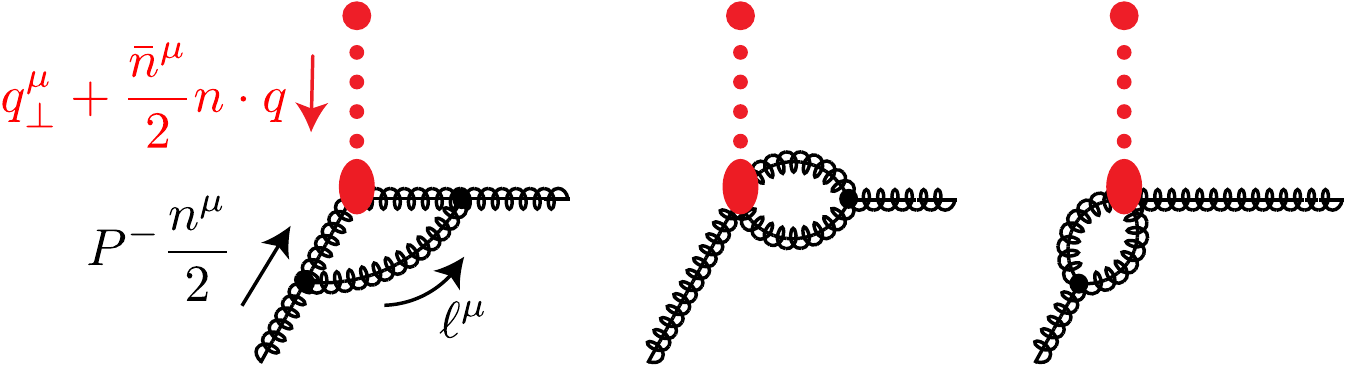}
    	\includegraphics[width=0.7\textwidth]{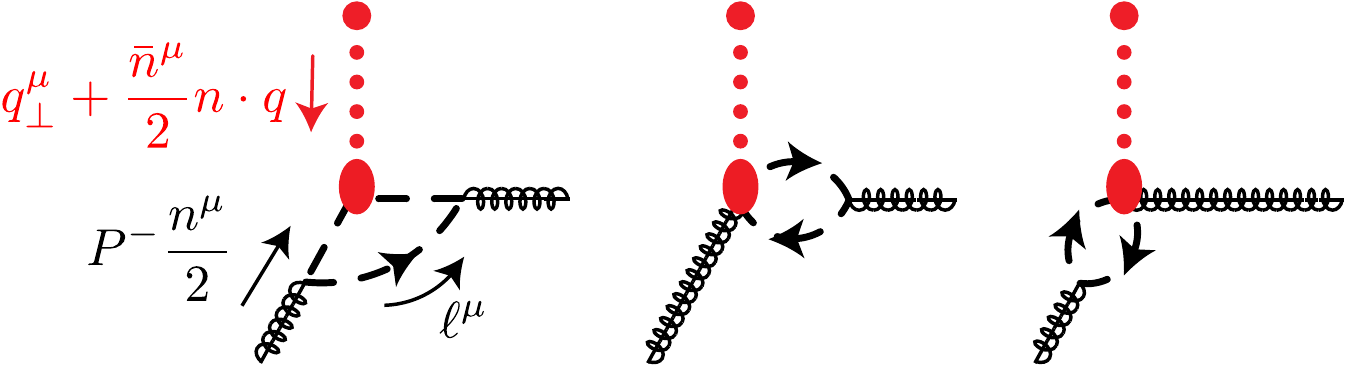}
	\vspace{-6pt}
	\caption{The virtual graphs for initial $n$-collinear gluon struck by an off-shell Glauber gluon.}
	\label{fig:gluonV}
\end{figure}
Next we evaluate the virtual graphs shown in \fig{gluonV}. Of these graphs, only the vertex correction graphs yield a non-trivial result such that
\begin{align}
	C_g^{(1), \text{virtual}}\Big(\frac{\nu}{\bn\cdot P}, \vec q^{\,2}_\perp\Big) &= 
	- \frac{	 \alpha_s C_A^2 }{\pi \vec q_\perp^{\,2}}\frac{\bn\cdot P}{\nu}    I_\eps\big[\vec q_\perp^{\,2}\big] \int_0^1\!\! \df z \: 
	\Big(P_{gg}^\eta (z)+\frac{n_fT_R}{2C_A}P_{q\bar{q}}(z)\Big) \big(z^{-2\eps} + (1-z)^{-2\eps}\big) .
\end{align}
Thus the result for the collinear function for an incoming gluon is given by
\begin{align}\label{eq:gluon1loop}
	&C_g^{(1) }\Big(\frac{\nu}{\bn\cdot P}, \vec q^{\,2}_\perp,\eps \Big) =   \frac{	 \alpha_s C_A^2 }{\pi \vec q_\perp^{\,2}}\frac{\bn\cdot P}{\nu} I_\eps\big[\vec q_\perp^{\,2}\big] \int_0^1 \df z \: 
	\Big(P_{gg}^\eta (z)+\frac{n_fT_R}{2C_A}P_{q\bar{q}}(z)\Big) \\
    	&\qquad=\frac{\alpha_s C_A}{\pi} C_g^{(0)}\Big(\frac{\nu}{\bn\cdot P}, \vec q_\perp^{\,2}\Big)(-2\pi) I_\eps\big[\vec q_\perp^{\,2}\big] \bigg(\frac{1}{\eta} + \log\Big(\frac{\nu}{P^-}\Big) + \frac{11}{12}-\frac{n_fT_R}{4C_A}\Big(1-\frac{1}{3(1-\epsilon)}\Big)\bigg) \, .\nn 
\end{align}

\subsubsection{BFKL Equation}
\label{sec:BFKL}
The rapidity log in the one-loop results in \eqs{quark1loop}{gluon1loop} is generated by the BFKL equation. In Ref.~\cite{Rothstein:2016bsq} the one-loop result soft function for forward $pp \ra pp$ scattering was derived and was shown to satisfy BFKL equation. This implies that by RG consistency of the effective theory, the collinear function must satisfy the same BFKL equation but with a relative factor of $-1/2$. The BFKL equation in $2-2\epsilon$ dimensions for our collinear function then reads
\begin{align}\label{eq:BFKL}
\nu\frac{d}{d\nu}C\Big(&\frac{\nu}{\bar{n}\cdot P},{q}_\perp,\eps \Big)=-C\Big(\frac{\nu}{\bar{n}\cdot P},{q}_\perp,\eps \Big)	\\
& -2\frac{\alpha_s C_A}{\pi} \iota^\eps\mu^{2\epsilon}\int\frac{d^{2-2\epsilon}{k}_\perp}{(2\pi)^{1-2\epsilon}}\Biggl\{\frac{C\Big(\frac{\nu}{\bar{n}\cdot P},\vec{k}_\perp,\eps \Big)}{(\vec{q}_\perp-\vec{k}_\perp)^2}-\frac{\vec{q}_\perp^{\,2}}{2\vec{k}_\perp^{\,2}(\vec{q}_\perp-\vec{k}_\perp)^{\,2}}C\Big(\frac{\nu}{\bar{n}\cdot P},{q}_\perp,\eps \Big)\Biggr\}\,.	\nn
\end{align}
Here the first term results from the classical scaling of the coefficient function $\sim 1/\nu$, and the second term involving the BFKL kernel describes the anomalous scaling. 

This renormalization group equation for $C$ could have been derived directly at the level of our ${\cal O}(\alpha_s)$ calculation of the collinear function in \secn{CollFunc}.
As a cross check we can confirm that  after subtracting the rapidity divergence, the coefficient of the associated logarithm in the rapidity renormalized collinear function agrees with the logarithm generated by the BFKL equation.  Simply plugging in the tree-level result of \eq{CnTree} into the right hand side of \eq{BFKL} gives:
\begin{align}
C_{\kappa\, \rm LL}\Big(\frac{\nu}{\bar{n}\cdot P},{q}_\perp,\eps\Big)&=C_\kappa^{(0)}\Big(\frac{\nu}{\bar{n}\cdot P},{q}_\perp\Big)\Bigg(1-\frac{\alpha_s C_A}{\pi} (2\pi)I_\eps\big[\vec q_\perp^{\,2}\big]\text{ln}\Big(\frac{\nu}{\bar{n}\cdot P}\Big) \Bigg)+ \cO(\as^2) \,.
\end{align}
where $I_\eps\big[\vec q_\perp^{\,2}\big]$ is given by \eq{Ieps}. The LL subscript signifies that this result at ${\cal O}(\alpha_s)$ does not include terms not predicted by BFKL. We find that it matches the rapidity logarithm found in the calculation of the collinear functions of Eqs. \eqref{eq:quark1loop} and \eqref{eq:gluon1loop}, verifying that the collinear sector  obeys the expected BFKL equation at  this order. In the next section we will perform resummation of small-$\x$ logarithms at LL, which only requires the tree level result of the collinear function.

%%%%%%%%%%%%%%%%%%%%%%%%%%%%%%%%%%%%%%%%%%%%%%%%%%%%%%%%%%%%%%%%%%%%%%%%%%%%%%%%
\section{BFKL and DGLAP Resummation}
\label{sec:resummation}
%%%%%%%%%%%%%%%%%%%%%%%%%%%%%%%%%%%%%%%%%%%%%%%%%%%%%%%%%%%%%%%%%%%%%%%%%%%%%%%%

Our strategy is to first perform small-$\x$ resummation of the structure function and then identify the resummed coefficient function and anomalous dimensions by performing the twist expansion. 
We first start from \eq{WmunuFinal} setting $\nu = \nu_S =\frac{\qsq}{n\cdot q} =x_b P^-$, which trivializes all rapidity logs in the soft function, such that
\begin{align}\label{eq:fact_again}
	\frac{1}{\x}	F^\kappa_p(\x , Q^2) = \int \df^{d-2} q_\perp^{\prime}	S_p\big(1, q_\perp, q_\perp',\eps\big)	C_\kappa \big(x_{b}, q_\perp',\eps\big)	 	\,	.
\end{align}
Naively we might imagine that the PDF, being a collinear object at a smaller invariant mass, would be entirely contained in the $C_\kappa$ function which involves the proton matrix element, whereas the vacuum matrix element $S_p$ would only account for finite process dependent pieces. However, as we saw above, the convolution between the soft and collinear functions generates IR divergences, and both have IR divergences at higher orders in loop-expansion, with some of them tied to BFKL evolution in $2-2\eps$ transverse dimensions.  (We have seen explicit additional IR divergence in the collinear function through our NLO calculation, though for the soft function, it may be the case that all IR divergences are tied to the convolution/BFKL logarithms.) Thus only after combining the two are we allowed to perform the twist expansion and identify the $1/\eps$ poles that are captured by the PDF. This is because the collinear and soft functions are at the same virtuality $\sim Q^2$, and while at finite $q_\perp'$ the tree level soft and collinear functions are IR finite, 
as $q_\perp' \to 0$ the integral in \eq{fact_again} enters the nonperturbative region $q_\perp'\sim \Lambda_{\rm QCD}$, 
inducing IR divergences, as seen from \eq{S2Leps}. 

%%%%%%%%%%%%%%%%%%%%%%%%%%%%%%%%%%%%%%%%%%%%%%%%%%%%%%%%%%%%%%%%%%%%%%%%%%%%%%%%

\subsection{Small-$x$ resummation in the EFT}

%%%%%%%%%%%%%%%%%%%%%%%%%%%%%%%%%%%%%%%%%%%%%%%%%%%%%%%%%%%%%%%%%%%%%%%%%%%%%%%%
\label{sec:collFuncLL}
We now discuss the resummation of the collinear function at leading logarithmic accuracy using the BFKL equation stated above. The resummation is best performed in Mellin space, where we Mellin-transform with respect to $\x$.
Taking the Mellin moment of \eq{fact_again},
\begin{align}
	\bar F^\kappa_p \big(N, Q^2\big)	=	 \int \df^{d-2} q_\perp^{\prime}	S_p\big(1, q_\perp, q_\perp',\eps\big)\bar 	C_\kappa \big(N, q_\perp',\eps\big)	\,	.
\end{align}
Likewise, having set $\nu=x P^-$ the Mellin transform of the BFKL equation in \eq{BFKL} gives
\begin{align}\label{eq:BFKLn0}
	\int_0^1 \df x 	\: x^{N-1}	\Big(x\frac{\df}{\df x}C_\kappa(x, q_\perp',\eps)\Big)	=	- \bar C(N, q_\perp',\eps)	-	\bar \alpha_s \iota^\eps K \otimes_\perp \bar C_\kappa \big(N, q_\perp',\eps\big)	\,	,	
\end{align}
where we have defined the BFKL kernel acting on a function $f(q_\perp)$ as 
\begin{align}
	\big[K\otimes_\perp f\big] (q_\perp)	\equiv  \mu^{2\epsilon}\int\frac{d^{2-2\epsilon}\vec{k}_\perp}{(2\pi)^{1-2\epsilon}}\Biggl\{\frac{2f(k_\perp)}{(\vec{q}_\perp-\vec{k}_\perp)^2}-\frac{\vec{q}_\perp^{\,2}}{\vec{k}_\perp^{\,2}(\vec{q}_\perp-\vec{k}_\perp)^{\,2}}f(q_\perp)\Biggr\}	\,	,
\end{align}
and 
\begin{align}\label{eq:asbar}
	\bar \alpha_s \equiv \frac{\alpha_sC_A}{\pi}	\,	.
\end{align}
Next, using integration by parts the left hand side of \eq{BFKLn0} becomes
\begin{align} \label{eq:simLHS}
	\int_0^1 \df x \: x^{N-1}\bigg( x\frac{\df}{\df x} C_\kappa\big(x, q_\perp',\eps\big)\bigg)	&=	-N \bar{C}_\kappa(N,q_\perp')  
	+ \Big[x^{N} C_\kappa\big(x,q_\perp'\big)\Big]_0^{1}	
	\nn\\
	&= -N \bar{C}_\kappa(N,q_\perp',\eps)  	+ C_\kappa^{(0)}	(1, q_\perp')\, .
\end{align}
Since $C_\kappa(x,q_\perp') \sim\frac{1}{x} \alpha_s^\ell \ln^{(\ell-1)} x$, the boundary condition at $x =0$ vanishes for $N \ra 1^+$ due to the prefactor $x^N\to 0$. 
For $x= 1$ all the logarithmic terms vanish and at leading log accuracy we are left with the tree-level result stated above in \eq{CnTree}:
\begin{align}
	C_{\kappa}^{(0)}(x = 1, q_\perp')	&=	\frac{c_\kappa}{\pi \qpsq}	\,	,
\end{align}
where $c_\kappa = C_F,C_A$. Hence, combining \eqs{BFKLn0}{simLHS} we have
\begin{align}\label{eq:BFKL_Mellin}
	\bar C_\kappa\big(N, q_\perp',\eps\big)
	&=\frac{c_\kappa }{(N-1)\pi\qpsq} +\frac{\bar \alpha_s \iota^\eps}{(N-1)}K\otimes_\perp \bar C_\kappa\big(N,q_\perp',\eps\big)	\,	,
\end{align}
For a more compact presentation, below we will shift our Mellin moment variable to $n=N-1$, as in \secn{logx},  writing $\bar C_\kappa\big(N, q_\perp'\big)=\bar C_\kappa\big(n, q_\perp'\big)$. 
As noted in \Ref{Catani:1994sq}, the BFKL kernel $K$ is  no longer scale invariant in $d=4-2\eps$ dimensions, and hence it is not straightforward to solve for $\bar C_\kappa(n, q_\perp^{\prime})$ by looking for eigen-functions of $K$. As mentioned above in \secn{CH}, in \Ref{Catani:1994sq} a similar equation was setup for their gluon Green's function $\cF_g^{(0)}$ (with different boundary conditions) 
and solved by setting up  equations that iteratively relate results at each order in $\alpha_s$.  Following the same approach, we now derive the LL solution of our collinear function.
 We first note that the BFKL kernel acts on power law test functions as
\begin{align}
	\iota^\eps	K \otimes_\perp \frac{1}{k_\perp^{2}}	\bigg(\frac{ k_\perp^{\,2}}{\mu^2}\bigg)^\gamma =	 \frac{e^{\eps \gamma_E}}{\Gamma(1-\eps)}\frac{1}{k_\perp^{2}}	\bigg(\frac{k_\perp^{2}}{\mu^2}\bigg)^{\gamma-\eps}	I(\gamma; -\eps)
\end{align}
with
\begin{align}
	I(\gamma; \eps) &\equiv  \Gamma(\eps)\Gamma(1+\eps)
	\bigg[
	\frac{\Gamma(\gamma)\Gamma(1-\gamma)}{\Gamma(\eps+\gamma) \Gamma(1 + \eps - \gamma)} - \frac{\Gamma(1+\eps)\Gamma(1-\eps)}{\Gamma(1+2\eps)}
	\bigg] 	\,	.
\end{align}
Hence, iteratively solving \eq{BFKL_Mellin} starting from the tree level result with $\gamma = 0$, we find 
\begin{align}\label{eq:CnBFKL}
	\bar C_{\kappa,\rm LL} (n ,q_\perp',\eps)	=\frac{1}{n}	\frac{c_\kappa}{\pi \qpsq} \sum_{\ell = 0}^\infty c_{\ell+1}(\eps)\bigg(\frac{\bar \alpha_s}{n}\frac{e^{ \eps\gamma_E}}{\Gamma(1-\eps)}\Big(\frac{\qpsq}{\mu^2}\Big)^{-\eps}\bigg)^{\ell}	\,	,
\end{align}
where
the coefficients $c_k(\eps)$ are given by
\begin{align}\label{eq:ckDef}
	c_1(\eps) = 1\, , \qquad c_{\ell+1}(\eps) = c_\ell(\eps) I(- \ell\eps; -\eps) \, ,\qquad  \ell\geq 1 \, . 
\end{align}

From the above LL result for the collinear function we can derive the small-$\x$ resummed strucutre function using \eq{Gamma_form_factorization} for finite $\eps$, such that
\begin{align}\label{eq:Fexp}
	\bar F_{p,\rm LL}^{\kappa}	(n,Q^2)=\frac{c_\kappa}{n\pi}\Big(\frac{\qsq}{\mu^2}\Big)^{-2\eps}	 	\sum_{\ell = 0}^\infty d_{p,\ell +1}(\eps)\bigg(\frac{\bas}{n}	\frac{e^{ \eps\gamma_E}}{\Gamma(1-\eps)}\Big(\frac{\vec q^{\,2}_\perp}{\mu^2}\Big)^{-\eps}\bigg)^{\ell}\,	,
\end{align}
where the coefficients are given by
\begin{align}
	d_{p,\ell + 1}(\eps)	\equiv c_{\ell + 1}(\eps)	\tilde S_p(1,-\ell\eps,\as,\eps)	\,	.
\end{align}

Having obtained the LL-resummed solution of the structure function, our next task is to isolate the IR-divergent terms and obtain results for the LL-resummed results for the coefficient function and anomalous dimensions.
Before we perform this calculation in detail, we first illustrate how the leading poles of the form $\big(\as/\eps\big)^\ell$ exponentiate in practice. For simplicity, let us set $\mu^2 = {\vec q}_\perp^{\,2}$ in \eq{Fexp} and consider the $F_L^g$ structure function. The coefficients $c_\ell$ in \eq{ckDef} for $\eps \ra 0$ behave as:
\begin{align}
	c_{\ell}(\eps)	= \frac{1}{\ell! }\Big(\frac{-1}{\eps}\Big)^{\ell}\Big(1 + \cO(\eps^2) \Big)\,	.
\end{align}
Combining the leading behavior of $\tilde S_L$ from \eq{S2Leps}, we have:
\begin{align}
	\frac{1}{n}\Big(\frac{\bas}{n}\Big)^\ell d_{\ell + 1}(\eps)	=	\frac{2\as n_f T_F}{3\pi}\bigg[ 	\frac{1}{(\ell+1)!}\Big(\frac{-1}{\eps}\frac{\bas}{n}\Big)^{\ell+1}	+ \cO(\eps^{-\ell})\bigg] 
 \,.
\end{align}
Hence, we find from \eq{Fexp} (recalling the definition of $\bas$ in \eq{asbar}):
\begin{align}\label{eq:FLLheuristic}
	\bar F_{L,\rm LL}^g (n)	+ \frac{2\as n_f T_F}{3\pi}	&=	 \frac{2\as n_f T_F}{3\pi}	\Bigg[\sum_{\ell = 0}^\infty\frac{1}{\ell!} \Big(\frac{-1}{\eps}\frac{\bas}{n}\Big)^\ell \big(1+\cO(\eps)\big) \Bigg]	\nn\\
	&=	\frac{2\as n_f T_F}{3\pi}\exp\bigg(-\frac{1}{\eps}\frac{\bas}{n}\bigg)\bigg(1+\cO\Big(\frac{\bas}{n}\Big)\bigg)+ \cO\bigg(\frac{1}{\eps}\Big(\frac{\bas}{n}\Big)^2\bigg)	\,	.
\end{align}
We see that in order to exponentiate the leading poles we necessarily needed to add $\frac{2\as n_f T_F}{3\pi}$, which is in fact a power suppressed contribution to the leading power $\bar F_{L,\rm LL}^g$ result. 
This points to a key subtlety when matching to twist expansion: we must take into account some of the power suppressed terms (in $\x$) when connecting the small-$\x$ resummed result with the $\Lambda_{\rm QCD}^2 \ll Q^2$  factorization. Because  of this  we will have to be careful when separating IR divergences from the LL small-$\x$ resummed structure functions in \eq{Fexp}. 

We pause to make some further remarks concerning power suppressed pieces in  Mellin-space. We have seen that in the $\x$-space, the EFT captures all the terms in the perturbative expansion of the cross section that scale as $1/x \ln^{k}(x)$. In the Mellin-space, we have effectively performed a Laurent expansion about $n = 0$ and all the polynomial pieces in $n$ are localized at $\delta(1-x)$ in the $x$-space and expanded away. This is reflected in our result in \eq{Fexp} that manifestly scales as $1/n$. In \eq{FLLheuristic}, the additional $\cO(\as)$ piece we added is given by Mellin transform of the ${\cal O}(\as)$ coefficient function $\tilde h_{L,1}^{(g)}(1,N)=\tilde h_{L,1}^{(g)}(1,n)$ defined through \eq{Hexpansion}:
	\begin{align}
		\tilde h_{L,1}^{(g)}(x)	=	4 n_f T_F x(1-x)	\, , \qquad \tilde h_{L,1}^{(g)}(n)	=	\frac{2 n_f T_F}{3} \frac{6}{6+5 n + n^2}	\,,
	\end{align}
and taking $n\to 0$.
However, we also notice that in the twist-expansion, the coefficient function also contains $\delta (1-x)$ terms, for example $\tilde h_{2,q}^{(0)}(x) = 2 n_f \delta (1-x)$ which corresponds to the tree-level diagram of the ``direct piece'' shown in \fig{direct_term}, and contributes to the $\delta_{p,2}\delta_{\kappa,2}$ term in \eq{Hexpansion}. While such terms contribute to $\cO(n^0)$ pieces in Mellin space, they have no well-defined expansion about $x = 0$. In other words, $\delta(1-x)$ pieces cannot be captured in the small-$\x$ EFT at any finite order in the power expansion. We denote such pieces as being \textit{irregular} in the small-$\x$ power counting. Nevertheless, as seen above for the $F_L^{(g)}$ case, they will be relevant for the DGLAP resummation analysis discussed in the next section. For simplicity, we will continue to refer to these terms as power suppressed or higher power pieces.

We note that despite common features between our results and the approach of Ref.~\cite{Catani:1994sq}, such as the use of the BFKL equation in $d$-dimensions, there are also some significant differences.
Firstly, the result in \eq{Fexp} is strictly leading power, and accordingly starts at $\cO(\as^2)$, which is made manifest by the soft functions $\tilde S_p$ which start at this order. 
When we carry out the twist expansion, the exponentiation of infrared divergences will necessitate the addition of formally power suppressed terms, and this power counting is
not manifest in the derivation of Ref.~\cite{Catani:1994sq} discussed in \secn{CH}.  
Secondly, although the LL collinear functions 
$\bar C_{g,{\rm LL}}$ and $\bar C_{q,{\rm LL}}$
in \eq{CnBFKL} have a similar expansion to that of quark and gluon 
channels of the gluon Green's functions $\cF_{q,g}^{(0)}$ in \eq{GluonGreen},
we have not needed to define an object analogous  to their quark Green's function in \eq{QuarkGreen} in our EFT, which they used for resummation of the $\gamma_{qg}$ anomalous dimension.
As we will see in more detail below,  in our approach the combination of our soft and collinear functions is sufficient to achieve resummation for all the components of the DGLAP anomalous dimension matrix.
These same soft and collinear functions also automatically incorporate scheme dependence, such as the constants present in the ${\overline {\rm MS}}$ scheme, again without the need for additional computations 
such as for the $R$ function in \eq{FLgCH}.

%%%%%%%%%%%%%%%%%%%%%%%%%%%%%%%%%%%%%%%%%%%%%%%%%%%%%%%%%%%%%%%%%%%%%%%%%%%%%%%%

\subsection{DGLAP resummation}

%%%%%%%%%%%%%%%%%%%%%%%%%%%%%%%%%%%%%%%%%%%%%%%%%%%%%%%%%%%%%%%%%%%%%%%%%%%%%%%%

We are now in the position to obtain leading logarithmic results for the coefficient functions and the PDF anomalous dimensions. 
In the twist expansion, using perturbation theory with dimensional regularization, the bare structure function factorizes as
\begin{align}\label{eq:FLsplit}
	\bar F_{p}^\kappa (n, Q^2)	=\sum_{\kappa'} 	\bar H_p^{(\kappa')}\Big(n, \frac{Q^2}{\mu^2},\as(\mu^2)\Big)	\bar \Gamma_{\kappa' \kappa} \big(\alpha_s(\mu^2), n\big) + \cO\bigg(\frac{\Lambda_{\rm QCD}^2}{Q^2}\bigg)\,	,
\end{align}
where the $n$-space transition function for parton $\kappa \ra \kappa'$ defined in \eq{TransitionFunction} captures the infra-red divergences of the perturbative calculation. In the fixed coupling approximation we have 
\begin{align}\label{eq:GammaijDef}
	\bar \Gamma_{\kappa'\kappa}	\big(\alpha_s(\mu^2), n\big)=	{\rm P}	\exp \bigg(-\frac{1}{\eps}\int_0^{\as(\mu^2)} \frac{\df \alpha}{\alpha}\mb \gamma^s(\alpha,n)\bigg)_{\kappa'\kappa}	\,	,
\end{align}
and $\bar \Gamma_{\kappa' \kappa}$ satisfies
\begin{align}\label{eq:AnomDimGamma}
	\mu^2\frac{\df}{\df \mu^2}	\bar \Gamma_{\kappa'\kappa}	\big(\alpha_s(\mu^2), n\big)	=	\sum_{i} \gamma_{\kappa' i}\big(\as(\mu^2),n\big)\bar \Gamma_{i\kappa}	\big(\alpha_s(\mu^2),n\big)	\,	.
\end{align}

In writing \eq{FLsplit} we have not performed an expansion in the small-$\x$ limit, and the coefficient function $\bar H_p^{(\kappa)}$ as well as the transition functions contain power suppressed terms which appear as poles at $n = -1, -2, \ldots$ in the Mellin space, which become polynomial upon Laurent expansion about $n=0$. While we do not resum towers of these power suppressed terms,  their product with leading power terms in Mellin space remains leading power.\footnote{This can be easily seen by writing such product terms with partial fractions. For example $\frac{1}{n(n+1)} = \frac{1}{n} - \frac{1}{n+1}$, which upon inverse Mellin transform remains leading power.} Hence, we must continue to include all the power suppressed terms truncating to a fixed-order
in the $\alpha_s$ expansion that is necessary for a given logarithmic accuracy, when we wish to resum logarithms of $Q^2$ over $\Lambda_{\rm QCD}^2$.

To obtain leading power results in the small-$\x$ limit we necessarily require intermediate off-shell Glauber gluons between the incoming parton and the quark struck by the photon. Thus we anticipate $\gamma_{gg}\sim \gamma_{gq}\sim \cO\Big(\big(\frac{\as}{n}\big)^\ell\Big)$ for $\ell \geq 1$. This argument also implies that $\gamma_{qg}\sim \gamma_{qq}\sim \cO\Big(\as\big(\frac{\as}{n}\big)^{\ell-1}\Big)$ for $\ell \geq 1$.  Then if we take the $\bar \Gamma_{\kappa\kappa'}$ to have the same log counting as the corresponding anomalous dimensions, we have
\begin{align}\label{eq:LL_Gamma}
	&\bar \Gamma_{gg}	=	1+\cO\bigg(\Big(\frac{\as}{n}\Big)^\ell\bigg)	\,	,&
	&\bar \Gamma_{gq}	=	\cO\bigg(\Big(\frac{\as}{n}\Big)^\ell\bigg)	\,	,&	\\
	&\bar \Gamma_{qg}	=	\cO\bigg(\as\Big(\frac{\as}{n}\Big)^{\ell-1}\bigg)	\,	,&
	&\bar \Gamma_{qq}	=	1+	\cO\bigg(\as\Big(\frac{\as}{n}\Big)^{\ell-1}\bigg)	\,	,&	\nn 	
\end{align}
where $\ell \geq 1$. We have explicitly indicated the logarithmic counting of the transition functions and we will self-consistently confirm this behavior in the small-$\x$ expansion below.
We do so by demanding consistency between the twist factorization of infra-red divergences and the small-$\x$ resummation. For  $\bar \Gamma_{\kappa'\kappa}$ with $\kappa\neq \kappa'$ we have made use of the fact that mixing  makes them start at  $\cO(\as)$. The renormalization group equations are then given by
\begin{align}
	  \mu^2\frac{\df}{\df \mu^2}\bar\Gamma_{gg}&=\gamma_{gg}\bar\Gamma_{gg}	+	\cO\bigg(\as\Big(\frac{\as}{n}\Big)^\ell\bigg)\,,\\  
	   \mu^2\frac{\df}{\df \mu^2}\bar\Gamma_{gq}&	=\gamma_{gg}\bar\Gamma_{gq}	+	\cO\bigg(\as\Big(\frac{\as}{n}\Big)^\ell\bigg)	\,,	\nn	\\
	\mu^2  \frac{\df}{\df \mu^2}	\bar \Gamma_{qg}	&=	\gamma_{qg}\bar\Gamma_{gg}	+	\cO\bigg(\as^2\Big(\frac{\as}{n}\Big)^{\ell-1}\bigg)\,,	\nn\\  
	  \mu^2\frac{\df}{\df \mu^2}\bar\Gamma_{qq}&=\gamma_{qq} 	+\gamma_{qg}\bar\Gamma_{gq}	+		\cO\bigg(\as^2\Big(\frac{\as}{n}\Big)^{\ell-1}\bigg)\,.\nn
\end{align} 
Recalling that $\mu^2\frac{\df}{\df \mu^2}\as=-\epsilon\as$, these equations have solutions:
\begin{align}
\label{eq:LL_gg}
\bar\Gamma_{gg}\big(n,\as(\mu^2)\big)&=\exp\Big(-\frac{1}{\eps}\int_{0}^{\as(\mu^2)}\frac{\df \alpha}{\alpha}\gamma_{gg}(n,\alpha)\Big)+ \cO\bigg(\as\Big(\frac{\as}{n}\Big)^\ell\bigg)
\,, \\
\bar\Gamma_{gq}\big(n,\as(\mu^2)\big)&=c\bar\Gamma_{gg}\big(n,\as(\mu^2)\big)+\text{const}+ \cO\bigg(\as\Big(\frac{\as}{n}\Big)^\ell\bigg)
\,,  \\
\label{eq:LL_qg}
\bar\Gamma_{qg}\big(n,\as(\mu^2)\big)&=-\frac{1}{\eps}\int_{0}^{\as(\mu^2)}\frac{\df \alpha}{\alpha}\gamma_{qg}(n,\alpha)\bar\Gamma_{gg}\big(n,\alpha\big)+ \cO\bigg(\as^2\Big(\frac{\as}{n}\Big)^{\ell-1}\bigg)
\,,\\
\bar \Gamma_{qq} \big(n,\as(\mu^2)\big)&=	1 - \frac{1}{\eps}\int_0^{\as(\mu^2)}	\frac{\df \alpha}{\alpha}	\big[\gamma_{qg}(\alpha)\bar \Gamma_{gq}(n,\alpha)	+	\gamma_{qq}(\alpha)\big]	+ \cO\bigg(\as^2\Big(\frac{\as}{n}\Big)^{\ell-1}\bigg)	
\,	.	
\end{align}
These results have two undetermined constants, $c$ and ``const''.
In the solution of $\bar \Gamma_{gq}$ the ``const'' term is independent of $\as$. Since from \eq{LL_Gamma} $\bar \Gamma_{gq}$  starts at $\cO(\as)$ due to mixing, ``const'' must be set to $-c$ to cancel the $\cO(1)$ piece in $c \bar \Gamma_{gg}$, such that
\begin{align}
	 \label{eq:LL_gq_ansatz}\bar\Gamma_{gq}\big(n,\as(\mu^2)\big)&=c\Big(\bar\Gamma_{gg}\big(n,\as(\mu^2)\big)-1\Big)\,,		\qquad 
	 \gamma_{gq}=	c	\gamma_{gg}\,	.
\end{align} 
Furthermore, in terms of $\bar \Gamma_{gg}$, the solution of $\bar \Gamma_{qq}$ is then given by
\begin{align}\label{eq:LL_qq}
	\bar \Gamma_{qq}\big(n, \as(\mu^2)\big)=	1	+ \frac{1}{\eps}	\int_0^{\as}\frac{\df \alpha}{\alpha}	\big(c\gamma_{qg} (\alpha)-\gamma_{qq}(\alpha)\big)	-\frac{c}{\eps}	\int_0^{\as}\frac{\df \alpha}{\alpha}	\: \gamma_{qg} (\alpha)\bar \Gamma_{gg}	(n,\alpha)	\,.
\end{align}
We will make use of these relations below in deriving the resummed results for the coefficient functions, and  will at the same time determine $c$.

\subsubsection{DGLAP resummation of $F_L$}
We first consider the case of $p = L$ and $\kappa = g$. 
As discussed near \eq{Hexpansion}, both $\bar H_L^{(q)}$ and $\bar H_L^{(g)}$ start at $\cO(\as)$, though this is a power suppressed contribution. Using the results of \eq{LL_gg} and \eq{LL_Gamma}, we find: 
\begin{align}\label{eq:FLg}
	\bar F_L^g\big(n\big)\equiv 
    \bar F_L^g\big(n,Q^2=\mu^2\big)	=	\bar H_L^{(g)}\Big(n, \frac{Q^2}{\mu^2}=1,\as\Big)	\bar \Gamma_{gg} \big(\alpha_s, n\big)	
\,.
\end{align}
To proceed further, we express the anomalous dimension $\gamma_{gg}$ and the coefficient function as a power series in $\as$ and $\eps$: 
\begin{align}
	&\bar H_L^{(g)}\Big(n, \frac{Q^2}{\mu^2} =1,\as\Big)	=\frac{\as}{\pi}\sum_{k=0}^\infty \eps^k	h_{L,g}^{(0,k)}	+	\frac{\as}{\pi}\sum_{\ell = 1}^{\infty}	\Big(\frac{\as}{\pi n}\Big)^\ell \sum_{k=0}^\infty \eps^k	h_{L,g}^{(\ell,k)}	\, ,\\
	&\gamma_{gg}=	\sum_{\ell =1}^\infty \gamma_{gg,\ell-1}\Big(\frac{\as C_A}{\pi n}\Big)^\ell \,	.
\end{align}
Finally, in the small-$\x$ power counting, we can write the structure function as:
\begin{align}%\label{eq:FLg}
	\bar F_L^{g}\big(n\big)	&=	\bar F_{L,\nlp}^{g}	+	\bar F_{L,\rm LL}^{g}(n)	\,	,
	\\
	\bar F_{L,\nlp}^{g}	&=	\frac{\as}{\pi}  \sum_{k=-1 }^\infty \eps^k	f_{L,g}^{(k)}		\,	.\nn
\end{align}
Here the subscript `$\nlp$' denotes higher power terms in the small-$\x$ power expansion that are required for consistency with the twist expansion. As mentioned above, these terms also include irregular $\delta(1-x)$ pieces.
In the power suppressed part of the $\bar F_L^g$ structure function, we only need to retain the $\cO(\as)$ term for leading logarithmic resummation, and we can take the limit $n\rightarrow 0$ as all poles in $n$ have been removed. We have allowed for the possibility that this term can be IR divergent. By sequentially comparing the coefficients of $(\as/\eps)^\ell$, $\as(\as/\eps)^\ell, \ldots$ terms using \eqss{Fexp}{LL_gg}{FLg}, we can straightforwardly solve for the unknown $h_{L,g}^{(\ell,k)}$, $f_{L,g}^{(k)}$ and $\gamma_{gg,\ell-1}$ terms, such that
\begin{align}\label{eq:FLgsol}
	\gamma_{gg} 	&=	\frac{\bas}{n}	+ 2 \zeta_3 \Big(\frac{\bas}{n}\Big)^4	+ \ldots	\\
	\bar H_L^{(g)}	&=	\frac{2\as n_f T_F}{3\pi}	\bigg(1 -\frac{1}{3}\frac{\bas}{n}	+ \Big(\frac{34}{9}-\zeta_2\Big) \Big(\frac{\bas}{n}\Big)^2
	+ \Big(-\frac{40}{27}	+ \frac{\pi^2}{18}	+ \frac{8}{3}\zeta_3\Big)	\Big(\frac{\bas}{n}\Big)^3	+ \ldots 
	\bigg)	\,	,	\nn\\
	\bar F_{L,\nlp}^{g}	&=	\frac{2\as n_f T_F}{3\pi}	\bigg(1+3\eps + \Big(6 - \frac{1}{2}\zeta_2\Big)\eps^2 + \Big(12 - \frac{\pi^2}{4}	-	\frac{7}{3}\zeta_3\Big)\eps^3 + \ldots \bigg)	\,	.\nn
\end{align}
For simplicity we have only shown the first few terms in the infinite series.
In the second series we have set $\eps = 0$ for simplicity, and the $\cO(\as)$ term of $\bar H_L^{(g)}$ is found to be the same as $F_{L,\nlp}^{(g)}$. 
Our results for the resummation in both $\gamma_{gg}$ and $\bar H_L^{(g)}$ agree with those of \Ref{Catani:1994sq}, 
including higher order terms that are not shown.
Interestingly, we see that consistency with the twist expansion automatically constrains the power suppressed non-singular pieces in the structure function and the coefficient function, although they are not determined ``mechanically'' within the calculation of the Feynman diagrams and rapidity resummation of the EFT for small-$\x$. Furthermore, it is important to retain the higher order terms in $\eps$ in $\bar F_{L,\rm HP}^{(g)}$ in \eq{FLgsol} as they determine the higher order terms in $\as/n$ in the coefficient function $\bar H_L^{(g)}$.

We next consider the quark channel for $\bar F_L^q(n,Q^2=\mu^2)$ structure function. As in \eq{FLg}, we can write:
\begin{align}\label{eq:FLq0}
	\bar F_L^q	&=		\bar F_{L,\nlp}^{q} 	+ \bar F_{L,\rm LL}^{q} \,.
\end{align}
It is clear from the calculation of the collinear function at leading logarithmic accuracy, we must have the relation:
\begin{align}\label{eq:casimir_ll_x}
\bar F_{L,\rm LL}^{q}=\frac{C_F}{C_A}\bar F_{L,\rm LL}^{g}\,.
\end{align}  
To proceed further, we simply need the factorization structure for $\bar F_L^q$:
\begin{align}\label{eq:FLq1}
	\bar F_L^q	= \bar H_L^{(q)}\bar\Gamma_{qq}	+	\bar H_L^{(g)}\bar	\Gamma_{gq}\,.	
\end{align}
A simple one loop calculation (e.g., \Ref{Moch:1999eb}) shows $\bar H_L^{(q)}\sim\cO\big(\as n^0\big)$, such that using \eq{LL_Gamma} we can set $\bar \Gamma_{qq} = 1$ at LL accuracy. Setting $\bar  F_L^g	=	\bar H_L^{(g)}\bar \Gamma_{gg}$, we find the relation
\begin{align}\label{eq:FLq2}
		\frac{1}{\bar H_L^{(g)}}\bigg[\Big(\bar H_L^{(q)}	-	\frac{C_F}{C_A}	\bar H_L^{(g)}\Big)	-	\Big(\bar F_{L,\nlp}^q	-	\frac{C_F}{C_A}F_{L,\nlp}^g\Big)\bigg] &= \frac{C_F}{C_A}\big(\bar\Gamma_{gg}-1\big)	-\bar \Gamma_{gq}		\\
		&=\Big(\frac{C_F}{C_A}-c\Big)\big(\bar\Gamma_{gg}-1\big)	\,.\nn
\end{align}
In the second line we plugged in our ansatz in \eq{LL_gq_ansatz}.
The result for $(\bar \Gamma_{gg}-1)$ on the right hand side is a series of $1/\eps$ poles and manifestly leading power, 
whereas the left hand side can at most have next-to-leading power IR divergences. Hence, up to power corrections, the right hand side must vanish.
Thus, 
\begin{align}\label{eq:Gammagq}
	c =\frac{C_F}{C_A}\,, \qquad \bar \Gamma_{gq}	=	\frac{C_F}{C_A}\big(\bar\Gamma_{gg}-1\big)	\,	.
\end{align}
Then from the left hand side  of \eq{FLq2} we find
\begin{align} \label{eq:HLsinglet}
	\bar H_L^{({\ps})}\equiv \bar H_L^{(q)}	-	\bar F_{L,\nlp}^{q}	=	\frac{C_F}{C_A}	\bigg(\bar H_L^{(g)}	-	\frac{2\as n_f T_F}{3\pi}\bigg)	\,	.
\end{align}
This combination  is the pure-singlet contribution to the coefficient function $\bar H_L^{(q)}$.
 The result in \eq{HLsinglet}  scales as $\cO\big(\as(\frac{\as}{n})^k\big)$. Since $\bar F_{L,\nlp}^q$ has no leading small-$\x$ pieces, it must be equal to the $\cO(\as)$ power suppressed part of the $\bar H_L^{(q)}$ coefficient function, and hence must be finite, which is in line with our earlier assumption.
Once again, our result for the resummed coefficient $\bar H_L^{(\ps)}$ agrees with \Ref{Catani:1994sq}.

%===================
\subsubsection{DGLAP resummation of $F_2$}
%===================

We now turn to the case of $p = 2$ and $\kappa = g$.
Here, unlike for $p=L$, first term in the $\as$ expansion gives $\bar H_2^{(q)} = 2 n_f$ such that the $\bar \Gamma_{qg}$ term can no longer be ignored relative to the $\bar \Gamma_{gg}$ term in \eq{FLsplit}. On the other hand, we can ignore the higher order corrections to $\bar H_2^{(q)}$ for LL resummation as they would result in terms $\cO\big(\as^2(\frac{\as}{n})^k\big)$. Therefore we can write
\begin{align}\label{eq:F2gtwist}
	\bar F_2^{g} =	2n_f \bar \Gamma_{qg}	+ \bar H_2^{(g)}	\bar \Gamma_{gg}	\,	.
\end{align}
We make use of \eq{LL_qg}, and as before we must include $\cO(\as)$ power suppressed terms in the $\bar F_2^g$ structure function that are not directly predicted by the small-$\x$ effective theory. 
We can do this by introducing a series of unknown coefficients through an $\epsilon$ expansion,
\begin{align}\label{eq:F2gsmallx}
	\bar F_2^{g} 	=	\bar F_{2,\nlp}^g	+	\bar F_{2,\rm LL}^g	\,	,\qquad
	\bar F_{2,\nlp}^{g}	=	\frac{\as}{\pi}  \sum_{k=-1 }^\infty \eps^k	f_{2,g}^{(k)}		\,	.
\end{align}
where the term $\bar F_{2,\rm LL}^g$ is the LL result in \eq{Fexp} for $p = 2$.
Hence, we have the following unknown series:
\begin{align}
	\bar H_2^{g}\Big(n, \frac{Q^2}{\mu^2} = 1,\as\Big)	&=\frac{\as}{\pi}\sum_{k=0}^\infty \eps^k	h_{2,g}^{(0,k)}	+\frac{\as}{\pi}\sum_{\ell = 1}^{\infty}	\Big(\frac{\as}{\pi n}\Big)^\ell \sum_{k=0}^\infty \eps^k	h_{2,g}^{(\ell,k)}	\, , \\
	\gamma_{qg}&=	\sum_{\ell =1}^\infty \gamma_{qg,\ell-1}\Big(\frac{\as}{\pi}\Big)^\ell \,	,\nn	\\	
	\bar F_{2,\nlp}^{g}	&=	\frac{\as}{\pi}	\sum_{k=-1}^\infty \eps^k f_{2,g}^{(k)}	\,	.\nn
\end{align}
As before, comparing the leading, next-to-leading and so on poles on both sides of \eqs{F2gtwist}{F2gsmallx} we find
\begin{align}\label{eq:F2gsol}
	\gamma_{qg}	&=	\frac{\as T_F}{3\pi}	\bigg(1	+	\frac{5}{3}\frac{\bas}{n}	+	\frac{14}{9}\Big(\frac{\bas}{n}\Big)^2	+	\Big(\frac{82}{81}	+ 2\zeta_3\Big)\Big(\frac{\bas}{n}\Big)^3	+ \ldots\bigg)	\,	,	\\
	\bar H_2^{(g)}	&=	\frac{\as n_f T_F}{3\pi}	\bigg(1 +\Big(\frac{43}{9}	-	2\zeta_2\Big)\frac{\bas}{n}	+	\Big(\frac{1234}{81}-\frac{13}{3}\zeta_2 + \frac{4}{3}\zeta_3\Big)\Big(\frac{\bas}{n}\Big)^3+\ldots\bigg)	\,	,	\nn	\\
	\bar F_{2,\nlp}^g	&=	\frac{\as n_f T_F}{3\pi}\bigg(-\frac{2}{\eps}	+1 +	(1+\zeta_2)\eps + \Big(1-\frac{1}{2}\zeta_2 +\frac{14}{3}\zeta_3\Big)\eps^2	+	\ldots\bigg)	\nn\,	.
\end{align}
Again we show only the first few terms in the series for simplicity. 
We find that the $\cO(\as)$ piece of $\bar H_2^{(g)}$ is the finite part of  $\bar F_{2,\nlp}^{g}$. Again, while consistency of the factorization of IR divergences in the twist expansion allows us to calculate these $\cO(\alpha_s)$ terms from the small-$\x$ resummation, they are not directly calculated from the small-$\x$ EFT. Only the terms involving explicit $(1/n)^k$ factors are predicted by the EFT.
Here we determined the resummed $\gamma_{qg}$ and $\bar H_2^{(g)}$ in the same manner used for $\gamma_{gg}$ and $H_L^{(g)}$, and they again agree with \Ref{Catani:1994sq}.

We now turn to the computation of the $\bar F_2^q(n,Q^2=\mu^2)$ structure function whose twist expansion reads
\begin{align}\label{eq:F2qtwist}
	\bar F_2^q	&=	\bar H_2^{(q)} \bar \Gamma_{qq}		+	\bar H_2^{(g)} \bar \Gamma_{gq}		\nn\\
	&=	\bar H_2^{(q)} + 2n_f \big(\bar \Gamma_{qq}	-1\big)	+ \frac{C_F}{C_A}\bar H_2^{(g)}\big(\bar \Gamma_{gg}-	1\big)+ \cO\bigg(\as^2\Big(\frac{\as}{n}\Big)^{\ell-1}\bigg) 
	\,,
\end{align}
where $\ell \geq1$.
Here we have made use of the lowest order term  $\bar H_2^{(q)} =	2n_f + \ldots$. Next, we note that the leading small-$\x$ terms obey Casimir scaling from \eq{Fexp}, such that
\begin{align}\label{eq:F2qsmallx}
	\bar F_2^q	&=	\bar F_{2,\nlp}^q	+ \frac{C_F}{C_A}\big(\bar F_2^g	-	\bar F_{2,\nlp}^g\big)	\\
	&=		\bar F_{2,\nlp}^q	+ \frac{C_F}{C_A}\bigg(-\frac{2n_f}{\eps}\int_0^{\as} \frac{\df \alpha}{\alpha}\gamma_{qg}(\alpha)\bar \Gamma_{gg}	+	\bar H_2^{(g)}\bar \Gamma_{gg}	-	\bar F_{2,\nlp}^g\bigg)	\,	,	\nn
\end{align}
where in the last line we used \eqs{F2gtwist}{LL_qg}. Thus, comparing \eqs{F2qtwist}{F2qsmallx} and using the result for $\bar \Gamma_{qq}$ in terms of $\bar \Gamma_{gg}$ in \eq{LL_qq} (with $c = \frac{C_F}{C_A})$ we find
\begin{align}
	\bar H_2^{(q)} 	=
	\bigg[	\bar F_{2,\nlp}^q	+\frac{2n_f}{\eps}	\int_0^{\as} \frac{\df \alpha}{\alpha}	\: \gamma_{qq}(\alpha)\bigg]
	+ \frac{C_F}{C_A}\bigg[\bar H_2^{(g)} -	\bar F_{2,\nlp}^g-\frac{2n_f}{\eps}\int \frac{\df \alpha}{\alpha}	\: \gamma_{qg}(\alpha)\bigg]	\,	.
\end{align}
From \eq{F2gsol} we find that $1/\eps$ pole cancels between the last two terms in the second square bracket. Since $\bar H_2^{(q)}$ on the left hand side is IR finite we conclude that the combination in the first term must also be finite. This term in the first square brackets is power suppressed in small-$\x$ power counting. 
Thus, we see that the pure-singlet part of the $\bar H_2^{(q)}$ coefficient function scaling as $\cO\big(\as(\frac{\as}{n})^\ell\big)$ is given by
\begin{align} \label{eq:H2singlet}
	\bar H_2^{(\ps)}	&\equiv	
	\bar H_2^{(q)} -\bigg[	\bar F_{2,\nlp}^q	+\frac{2n_f}{\eps}	\int_0^{\as} \frac{\df \alpha}{\alpha}	\: \gamma_{qq}(\alpha)\bigg]
    \nn	\\
	&=\frac{C_F}{C_A}\bigg[\bar H_2^{(g)} -	\bar F_{2,\nlp}^g-\frac{2n_f}{\eps}\int \frac{\df \alpha}{\alpha}	\: \gamma_{qg}(\alpha)\bigg]	\nn	\\
	&=	\frac{C_F}{C_A}\bigg(\bar H_2^{(g)}	-	\frac{\as n_f T_F}{3\pi}\bigg)	\,	.
\end{align}
Similarly, we find that the $\mu$ dependence of the pure-singlet coefficient must be described by
\begin{align}\label{eq:gammaqq}
	\gamma_{\ps}	=	\frac{C_F}{C_A}\bigg(\gamma_{qg}	-	\frac{\as T _F}{3\pi}\bigg)	\,	.
\end{align}
Once again our resummed results in \eqs{H2singlet}{gammaqq} agree with \Ref{Catani:1994sq}.

%===================
\subsubsection{Discussion}
% ===================
\label{sec:discuss}

It is interesting to compare our approach to the derivation of the LL-resummed results above with that used in \Ref{Catani:1994sq}. As reviewed in \secn{CH}, 
for the analysis of $F_L^{\kappa}$, \Ref{Catani:1994sq} first determined the DGLAP anomalous dimensions $\gamma_{gg}$ and $\gamma_{gq}$ from their gluon Green's function. The  resummed coefficient functions $\bar H_L^\kappa$ were determined after combining the gluon Green's function with a separate calculation of an $\cO(\as)$ off-shell cross section $h_L(\gamma)$ in \eq{S2Lh2L}, and a $\overline{\rm MS}$ scheme conversion factor $R$. In the analysis of $\gamma_{qg}$, \Ref{Catani:1994sq} however needed to consider a different, quark Green's function which also necessarily included the purely collinear $1/\eps$ pole, a higher power IR divergent term we saw in $\bar F^g_{2,\nlp}$ in \eq{F2gsol}. 
Finally, the approach of \Ref{Catani:1994sq} works only  at LL accuracy and at higher logarithmic accuracy Green's functions alone (unless suitably generalized) cannot be used to yield resummed anomalous dimensions. 

In contrast, by matching our small-$x$ resummed EFT results at the scale $Q$ directly onto partonic PDFs (encoding physics at the scale $\Lambda_{\rm QCD}$), we simultaneously determined both the DGLAP anomalous dimensions $\gamma_{ij}$ as well as the LL-resummed coefficient functions $\bar H_p^{(i)}$, while also tracking the appearance of leading power versus higher power pieces.  We did not need to define separate quark and gluon Green's functions (or analogous objects) to address differences in the treatment of leading twist factorization of $F_2$ versus $F_L$ structure functions, or calculate a distinct function $R$ to encode the $\overline{\rm MS}$ scheme dependence. 

In our approach, such differences arise from the additional collinear divergences in $F_2$, found in the soft function's double pole structure when $\gamma\sim\epsilon$ in \eq{Gamma_form_soft}, leading to \eq{S2Leps}. We saw that the $\eps$-dependence of the soft function contributes to the calculation of the coefficient function $\bar H_p^{(\kappa)}$ and the anomalous dimension $\mb \gamma^s$ (more specifically, the $\gamma_{qg}$ component) and also plays a role in fixing the scheme dependence. Though the small-$\x$ EFT  at leading power does not mechanically capture all the terms necessary for the factorization of the infra-red divergences in the twist expansion, we have found that at leading logarithmic accuracy, we could determine the missing terms ($\bar F_{p\, \nlp}^\kappa$) through consistency of the factorization process. Alternatively, it is a simple matter to calculate the $\cO\big(\as\big)$ contributions to the structure function, and take the Laurent expansion about $n=0$ to find these terms. 

It is worth noting that the consistency we find also implies the following intriguing relation, which maps the small-$\x$ soft function with the off-shell Glauber momentum to an effectively on-shell collinear gluon coupled to soft quarks:
\begin{align}
	\lim_{n\rightarrow 0}\bar F_{p\,\nlp}^g\Big|_{\cO(\as)}&= \lim_{\gamma\rightarrow 0}\frac{\gamma}{\alpha_s}\frac{\Gamma(1-\epsilon)}{e^{\eps\gamma_E}} \tilde{S}_p\Big(1,\gamma+\epsilon,\alpha_s,\epsilon\Big)\,,
\end{align}  
where $\bar F_{p\,\nlp}^g$ are define in \eqs{FLg}{F2gsmallx}. This relation was also at play when we found our soft function results for finite-$\gamma$ to be straightforwardly related to the power suppressed offshell pieces of \Ref{Catani:1994sq} in \eq{S2Lh2L}.

%%%%%%%%%%%%%%%%%%%%%%%%%%%%%%%%%%%%%%%%%%%%%%%%%%%%%%%%%%%%%%%%%%%%%%%%%%%%%%%%
\section{Conclusion and Future Outlook}
%%%%%%%%%%%%%%%%%%%%%%%%%%%%%%%%%%%%%%%%%%%%%%%%%%%%%%%%%%%%%%%%%%%%%%%%%%%%%%%%

\label{sec:conclusion}

We have shown how to construct from the SCET framework with Glauber interactions the resummation of the small-$\x$ scattering cross-section in DIS at leading logarithmic accuracy, reproducing the classic work by Catani and Hautmann in Ref.~\cite{Catani:1994sq}. The key feature of our calculation is that after the derivation of the factorized and individually gauge invariant soft and collinear functions, the perturbative computation of these functions are made straightforwardly from their operator definitions. In constrast, Ref.~\cite{Catani:1994sq} made use of off-shell cross-sections to interface the BFKL resummation to the electromagnetic probe current, and these cross-sections can only be guaranteed  to be gauge invariant to leading-order, making the extension to find  higher order resummed  corrections challenging. Further, we have by-passed extracting the DLGAP anomalous dimensions from the BFKL Green's functions, finding this step unnecessary in our setup. 
 We have also carried out the NLO calculation of the collinear-function, which served as a consistency check but only enters the resummation analysis at higher orders.  Given this result and the known two-loop BFKL equation, constructing the next-to-leading order resummation for DIS coefficient functions requires only calculating the loop corrections to the soft function, and considering possible contributions from multiple Glauber exchange. This would help bring the BFKL approach to similar levels of accuracy being pursued in the ``B-JIMWLK'' formalism, for instance in Refs.~\cite{Balitsky:2012bs,Balitsky:2013fea,Kovner:2014lca,Lappi:2016fmu,Beuf:2016wdz,Beuf:2017bpd,Roy:2019hwr,Bergabo:2022zhe}. 

Next, we note that calculations analogous to our one-loop collinear function were carried out for ``impact-factors'' in Ref.~\cite{Ciafaloni:1998hu}. We find that the results for quark and gluon collinear functions in \eqs{quark1loop}{gluon1loop} agree with the results for impact factors Ref.~\cite{Ciafaloni:1998hu}, up to constant terms proportional to the two-loop cusp anomalous dimensions. This is similar to other calculations with the SCET Glauber Lagrangian and its separation into soft and collinear contributions~\cite{Moult:2022lfy}. Our expectation is that such constants are naturally associated to the soft sector of the effective theory, which allows our collinear functions to be process independent. Furthermore, the calculation of impact factors in Ref.~\cite{Ciafaloni:1998hu} required a careful subtraction of the Green's function pieces from the cross section, inducing factorization-scheme dependencies. On the other hand, the computation of the factorized functions in our small-$\x$ factorization formula follow straightforwardly from their operator definitions which can be defined in a definite scheme from the start. The computation from operator definitions at higher orders will require care while treating zero-bin subtractions, however this will not induce any process or factorization scheme dependence.

Lastly, it will be interesting to examine more differential observables, particularly those also sensitive to Sudakov effects, like those found in end-point of the fragmentation spectrum or transverse-momentum dependent parton distribution functions, to see how the effective theory accomplishes the separation of small-$\x$ logarithms for such terms.

%%%%%%%%%%%%%%%%%%%%%%%%%%%%%%%%%%%%%%%
\section*{Acknowledgements}
%%%%%%%%%%%%%%%%%%%%%%%%%%%%%%%%%%%%%%%
\enlargethispage{20pt}
We would like to thank Ira Rothstein for many conversations and early collaboration. We also like to thank Simone Marzani for colloboration at early stages. D.N. would also like to thank Ian Moult for conservations regarding the use of the Glauber Lagrangian in SCET. D.N. was supported by the Department of Energy under Contract DE-AC52-06NA25396 at LANL and through the LANL/LDRD Program. AP acknowledges support from DESY (Hamburg, Germany), a member of the Helmholtz Association HGF. AP was previously a member of the Lancaster-Manchester-Sheffield Consortium for Fundamental Physics, which is supported by the UK Science and Technology Facilities Council (STFC) under grant number ST/T001038/1, and at University of Vienna was supported by FWF Austrian Science Fund under the Project No.~P28535-N27. IS was supported by the U.S.~Department of Energy, Office of Science, Office of Nuclear Physics, from DE-SC0011090 and by the Simons Foundation through the Investigator grant 327942.

\appendix

\section{Glauber Action In Position Space}
\label{app:GlauberAction}

Here we describe the steps leading to \eq{Glauber_operator_positions_space} since we find it useful to make use of the Glauber Lagrangian of SCET with more fields in position space, relative to the presentation in Ref~\cite{Rothstein:2016bsq}. The Glauber action $S_G$ for soft-collinear interactions is given by
\begin{align}\label{eq:OnsPos}
    S_G = \sum_{i,j} \int \df^4 x \:  {\cal O}_{ns}^{ij} (x) \,, 
    \qquad
    {\cal O}_{ns}^{ij}(x)  =8\pi \alpha_s e^{-\im  x\cdot {\cal P}} {\cal O}_n^{iA}(\tilde x) \frac{1}{{\cal P}_\perp^2} {\cal O}_{s}^{j_n A}(\tilde x) \, , 
\end{align}
where $i,j = q,g$. In \eq{OnsPos} the coordinates $\tilde x^\mu  = (x^+, x^- , 0_\perp)$ are conjugate to ${\cal O}(\lambda^2)$ residual momenta.  The operator is composed of $n$-collinear fields for the incoming proton and the intermediate soft fields (including Wilson lines made out of $n\cdot A_s$ fields), given by
\begin{align}\label{eq:CollinearBilinears}
    &{\cal O}_{n_i}^{qA} = \overline \chi_{n_i} \mathbf  T^A_i \frac{\bnslash_i}{2} \chi_{n_i} \, ,&
    &{\cal O}_{n_i}^{gA} = \frac{1}{2} {\cal B}_{n\perp\mu}^B (\im f^{ABC}) \frac{\bn_i}{2} \cdot ({\cal P} + {\cal P}^\dagger) {\cal B}_{n\perp}^{C\mu}  \, ,& \\
    &{\cal O}^{n_i, qA}_s = \overline \psi_S^{n_i}  \mathbf T^A_i \frac{\nslash}{2} \psi_S^{n_i} \, ,&
    &{\cal O}^{n_i, gA}_s=  \frac{1}{2} {\cal B}_{S\perp\mu}^{n_iB}  (\im f^{ABC})\frac{n_i}{2} \cdot ({\cal P} + {\cal P}^\dagger) {\cal B}_{S\perp}^{n_iC\mu} \, ,& \nn
\end{align}
where $\mathbf T^A$ corresponds to (anti-)fundamental representation for (anti-)quarks. 
Here, the label momentum operator ${\cal P}^\mu$ selects the ${\cal O}(1)$ and ${\cal O}(\lambda)$ momentum components such that
\begin{align}
    {\cal P}^\mu = \frac{n^\mu}{2} \big({\cal P}^- + {\cal P}_s^-\big) + \frac{\bn^\mu}{2}\big({\cal P}^+ + {\cal P}_s^{+}\big)+ {\cal P}_\perp^\mu \, , \qquad {\cal P}^\pm \sim \lambda^0 \, , \qquad {\cal P}_s^\pm \sim {\cal P}_\perp^\mu \sim \lambda \, .
\end{align}
We follow the standard convention where $\cP^{\mu\dagger}$ ($\cP^\mu$) gives a positive (negative) sign for conjugate fields.

The operators in \eq{OnsPos} are defined with a sum over all possible label momenta of order $\lambda$ in the power counting, where we can write:
\begin{align}\label{eq:label_sum_collinear}
{\cal O}_n^{iA}(\tilde x)&=\sum_{\bar{n}\cdot k_g}\int \frac{\df^2k_{g\perp}}{(2\pi)^2}\int \frac{\df^2\tilde p_r}{2(2\pi)^2} e^{i\tilde x\cdot \tilde p_r}[{\cal O}_{n,\bar{n}\cdot k_g}^{iA}(\tilde p_r,k_{g\perp})]
 \,,\\
{\cal O}_s^{j_n A}(\tilde x)&=\sum_{\bar{n}\cdot k_s,n\cdot k_s}\int \frac{\df^2k_{s\perp}}{(2\pi)^2}\int \frac{\df^2\tilde p_r'}{2(2\pi)^2} e^{i\tilde x\cdot \tilde p_r'}[{\cal O}_{s,\bar{n}\cdot k_s,n\cdot k_s}^{j_n A}(\tilde p_r',k_{s\perp})]
  \,.
\label{eq:label_sum_soft}
\end{align} 
We have also pulled out the Fourier transform over the residual momenta. The operators on the right hand side have definite ${\cal O}(\lambda)$ label-momentum and ${\cal O}(\lambda^2)$ residual-momentum injected into them, and the phases shown here as well as the $e^{-ix\cdot \cP}$ in \eq{OnsPos}, will induce momentum conservation.
We note that the collinear operator \emph{must} have zero total large label momentum:
\begin{align}
{\cal P}^-{\cal O}_n^{iA}(\tilde x)=0\,.
\end{align}  
The soft operator it is tied to in the Glauber action cannot inject ${\cal O}(\lambda^0)$ momenta in the power counting, as the soft sector has no ${\cal O}(\lambda^0)$ momenta. Thus in any graph that the collinear operator is inserted, the same collinear momentum flowing into the collinear bilinear must flow out on the other line it connects to. Thus the net large momentum label of the operator is zero. We have therefore suppressed the ${\cal O}(\lambda^0)$ label momentum which is always conserved by Glauber operators.  The collinear operator can only carry a momentum label of order $\lambda$, which we denote as $\bar{n}\cdot k_g$ and $k_{g\perp}$, where the 
$g$ subscript  indicates that the momenta are injected along the Glauber line.

Substituting these results into \eq{OnsPos}, acting with the label operators, we can combine the label sums with the integrals over the residual momenta to form a ``continuous'' label:
\begin{align}
\sum_{\bar{n}\cdot k_g}&\int \frac{\df \bn\cdot p_r}{2\pi}\rightarrow \int\frac{\df \bn\cdot k_g}{2\pi}\,,\\
\sum_{\bar{n}\cdot k_s',n\cdot k_s'}&\int \frac{\df^2 \tilde p_r'}{(2\pi)^2}\rightarrow \int\frac{\df^2\tilde k_s}{(2\pi)^2}\,. \nn
\end{align}  
Where we have shown the recombination for both the soft and the collinear (sub)-label sums. The final result is:
\begin{align}
  S_G &=8\pi \alpha_s \int \df^4 x \int\frac{\df^4k_s}{(2\pi)^4}\int \frac{\df^4k_g}{(2\pi)^4} e^{ix\cdot(k_s+k_g)}[{\cal O}_{n}^{iA}(k_g)]\frac{1}{k_{s\perp}^2}[{\cal O}_{s}^{j_n A}(k_s)]\,,\\
  k_s &=(\bar{n}\cdot k_s,n\cdot k_s, k_{s\perp})\sim \sqrt{s}(\lambda,\lambda,\lambda)\,, \nn \\
  k_g &=(\bar{n}\cdot k_g,n\cdot k_g, k_{g\perp})\sim \sqrt{s}(\lambda,\lambda^2,\lambda)\,. \nn
\end{align}  
Next we introduce position space representation of the operators, allowing us to perform the integral over the position $x$, and the $k_s$ integral, and relabel  $k_g\rightarrow q'$: 
\begin{align}
    S_G &= 8\pi \alpha_s\sum_{i,j,A}\int \df^4y \int d ^4z \int \df^4k_s\int \frac{\df^4k_g}{(2\pi)^4}\delta^{(4)}( k_s+k_{g}) e^{iy\cdot k_s+iz\cdot k_g}[{\cal O}_{n}^{iA}(z)]\frac{1}{k_{s\perp}^2}[{\cal O}_{s}^{j_n A}(y)]\,.\nn\\
    &=8\pi \alpha_s \sum_{i,j,A}\int \df^4y \int d ^4z \int \frac{\df^4q'}{(2\pi)^4}  \frac{e^{iz\cdot q'-i y\cdot q'}}{q_\perp^{\prime2}} {\cal O}_{n}^{iA}(z){\cal O}_{s}^{j_n A}(y)\,. 
\end{align}

Finally to transition this result to $d$ dimensions, we work with a dimensionless coupling $\alpha_s(\mu)$ and write the convolution in \eq{OnsPos} as 
\begin{align}
    S_G = \big(\iota \mu^2\big)^{\frac{4-d}{2}} 8\pi \alpha_s(\mu) \sum_{i,j}  \int \df^d x \: \, {\cal O}^{ij}_{ns}(x) \,, 
\end{align}
such that both collinear and soft operators, ${\cal O}_n^{iA}$ and ${\cal O}_s^{j_n A}$, have dimensions $d-1$. The $\overline {\rm MS}$ factor $\iota$ was defined in \eq{MSbar}.  We have dropped the coupling renormalization factor $Z_\alpha$ which is not needed for our calculations.

\bibliography{x1dis}

\end{document}